\documentclass[reprint,author-numerical,nofootinbib]{revtex4-2}
\usepackage[utf8]{inputenc}
\usepackage[english]{babel}
\usepackage{graphicx}
\usepackage{dcolumn}
\usepackage{bm}
\usepackage{amsmath}
\usepackage{amsfonts}
\usepackage{amssymb}

\def\tr{\mathop{\rm tr}\nolimits}

\def\dif{{\rm d}}

\newcommand{\be}{\begin{equation}}
\newcommand{\ee}{\end{equation}}

\begin{document}

\preprint{AIP/123-QED}

\title[Hydrodynamic approach to the Synge gas]{Hydrodynamic approach to the Synge gas}

\author{Salvador Mengual}
\affiliation{ Departament d'Astronomia i Astrof\'{\i}sica,
Universitat de Val\`encia, E-46100 Burjassot, Val\`encia, Spain.}
\author{Joan Josep Ferrando}
\altaffiliation[Also at ]{Observatori Astron\`omic, Universitat
de Val\`encia,  E-46980 Paterna, Val\`encia, Spain}
\email{joan.ferrando@uv.es.} \affiliation{ Departament d'Astronomia
i Astrof\'{\i}sica, Universitat
de Val\`encia, E-46100 Burjassot, Val\`encia, Spain.}
\author{Juan Antonio S\'aez}
\affiliation{ Departament de Matem\`atiques per a l'Economia i
l'Empresa, Universitat de Val\`encia, E-46022 Val\`encia, Spain.}

\date{\today}

\begin{abstract}
The necessary and sufficient conditions for a perfect energy tensor to describe the energy evolutions of a monoatomic relativistic Synge gas are obtained. Then, a Rainich-like theory for the Einstein-Synge solutions can be constructed. Equations of state approximating that of a Synge gas at low or at high temperatures, or in the entire domain of applicability, are analyzed from a hydrodynamic point of view.
\end{abstract} 
\pacs{04.20.-q, 04.20.Jb}

\maketitle

\section{\label{sec-intro}Introduction}

The macroscopic equation of state (EoS) for a simple monoatomic gas composed of non-degenerate relativistic particles can be derived from the microscopic kinetic theory developed by J\"uttner \cite{Juttner}. This relativistic gas was studied by Synge \cite{Synge}, and the perfect fluids obeying this equation of state are known as Synge fluids \cite{Rezzolla}. The relativistic Synge gas is the only relativistic fluid for which a macroscopic EoS has been obtained from microscopic kinetic theory  \cite{Rezzolla}.  

Many high-energy astrophysical scenarios, such as accretion flows, jet flows, gamma-ray bursts, and pulsar winds, involve relativistic flows (see \cite{Mignone} and references therein). Some of the earliest applications of such an equation of state were carried out by Chandrasekhar \cite{Chandra} and Bisnovatyi-Thorne \cite{Thorne} (see also \cite{Chavanis} and references therein), who used it in order to construct stellar models, and Krautter et al. \cite{Krautter}, who used it to study galactic jets. Synge EoS has also been used to study a wide range of other physically relevant phenomena: from stellar winds \cite{Meliani} to relativistic shocks \cite{Lanza, Mignone}, which are not only interesting to study astrophysical objects but also laboratory plasmas.

A relativistic two-dimensional hydrodynamic code incorporating the Synge EoS was developed by Scheck et al. \cite{Scheck-Aloy} and later used by Perucho and Mart\'{\i} \cite{Perucho_2007} to study the evolution of extragalactic jets. More recently, Choi and Wiita \cite{Choi} and Perucho et al. \cite{Perucho_2014} have developed three-dimensional codes which also make use of the Synge EoS. These codes have been applied repeatedly to study jets from active galactic nuclei \cite{Perucho_2019, Perucho_2021, Perucho_2022}. A relativistic gas has also been used to investigate the description of the early Universe and the role of dark matter \cite{Szekeres-Barnes, deBerredo, Silva-2019}. 

A Synge fluid is a specific solution of the fundamental system of relativistic hydrodynamics defined by the following elements:

(i) A divergence-free perfect energy tensor that describes the hydrodynamic evolution of the fluid:
\begin{equation}\label{nablaT}
\nabla \cdot T = 0 \, , \qquad T = (\rho + p)u\otimes u + pg \, .
\end{equation}
This condition gives four equations for the five {\em hydrodynamic quantities} $\{u,\rho,p\}$, namely, the {\em unit velocity} $u$ of the fluid, its {\em energy density} $\rho$ and its {\em pressure} $p$.

(ii) A set of {\em thermodynamic quantities} $\{n, s, \Theta\}$, the {\em rest-mass density} $n$, the {\em specific entropy} $s$ and the {\em temperature} $\Theta$, constrained by the usual thermodynamic laws  \cite{Eckart}. Namely, the conservation of matter, 
\begin{equation} \label{conservacio_massa}
\nabla \cdot (nu) =  0 \, ;
\end{equation}
and the {\em local thermal equilibrium relation}, which can be written as
\begin{equation} \label{re-termo}
 \Theta \dif s = \dif h - \frac{1}{n} \dif p \, ,  \qquad h \equiv \frac{\rho+p}{n} \, ,
\end{equation}
where $h$ is the {\em relativistic specific enthalpy}.

(iii) The macroscopic equations of state of a relativistic non-degenerate monoatomic gas \cite{Synge, Rezzolla}
\begin{equation}
p = k n \Theta,   \qquad    h = h(z) \equiv {K_3(z) \over K_2(z)}  , \quad z \equiv {1 \over k \Theta},  \label{synge}
\end{equation}
$K_n(z)$ being the second kind modified Bessel functions.

Note that points (i) and (ii) define the deterministic fundamental system of the perfect fluid hydrodynamics, ${\cal F} \equiv \{(\ref{nablaT}) (\ref{conservacio_massa}) (\ref{re-termo})\}$, which characterizes the evolution of any perfect fluid in local thermal equilibrium. The first equation of state in (\ref{synge}) constrains the fluid to be a generic ideal gas, and the second equation in (\ref{synge}) forces this ideal gas to be a Synge gas.

In \cite{Coll-Ferrando-termo} we showed (see also the more recent papers \cite{CFS-LTE, CFS-CC}) that the necessary and sufficient condition for a divergence-free energy tensor $T$ to represent the energy evolution of a thermodynamic perfect fluid in local thermal equilibrium is that its hydrodynamic quantities $\{u,\rho,p\}$ fulfill the {\em hydrodynamic sonic condition}:
\begin{equation}\label{HSC}
\hspace{-14mm} {\rm S} : \qquad \qquad  \dif \chi \wedge \dif\rho \wedge \dif p = 0 \, , \qquad \chi \equiv \frac{u(p)}{u(\rho)} \, ,    
\end{equation}
where, for a function $q(x^\alpha)$, $\dif q$ denotes its differential, $\dif q = \partial_\alpha q \dif x^{\alpha}$, and $u(q) = i(u) \dif q = u^\alpha \partial_\alpha q$, and where $\wedge$ denotes the exterior product (antisymetritzation of the tensorial product). Constraint (\ref{HSC}) indicates that the {\em indicatrix function} $\chi$ is a function of state, $\chi=\chi(\rho,p)$, which coincides with the square of the speed of sound, $c_s^2 = \chi(\rho,p)$.  

The above result states that if $T\equiv \{u,\rho,p\}$ is a solution to the {\em hydrodynamic flow system} ${\cal{H}} \equiv \{(\ref{nablaT}) (\ref{HSC})\}$, then a thermodynamic scheme $\{n, s, \Theta\}$ exists such that $\{u,\rho,p,n, s, \Theta\}$ is a solution of the fundamental system ${\cal F} \equiv \{(\ref{nablaT}) (\ref{conservacio_massa}) (\ref{re-termo})\}$. This means that the local thermal equilibrium condition admits a purely hydrodynamic characterization.

It is worth remarking that a specific fluid has a specific expression for the function of state $c_s^2 =\chi(\rho,p)$. In particular, when the fluid is a generic ideal gas (the first equation in (\ref{synge}) holds), the indicatrix function is $\chi = \chi(\pi) \not= \pi$, $\pi \equiv p/\rho$ \cite{CFS-LTE}. The first goal of this paper is to show that the second equation of state in (\ref{synge}) imposes a first order differential equation on $\chi(\pi)$. This leads to a purely hydrodynamic characterization of a Synge gas. 

The hydrodynamic approach to the thermodynamic perfect fluid solutions offered by the sonic condition (\ref{HSC}), and to the ideal gas solutions offered by the {\em ideal sonic condition}, $\chi = \chi(\pi)$, has enabled us to analyze the physical meaning of significant families of perfect fluid solutions to the Einstein equations \cite{CFS-CC, CF-Stephani, CFS-CIG, FS-SS, CFS-PSS, CFS-RSS, FM-Tmodels, FM-Tmodels2, MF-LT}.

Similarly, the current hydrodynamic approach to the Synge gas can be useful for studying both test solutions and self-gravitating systems. 
Moreover, it enables us to analyze whether several equations of state, which approach that of a Synge gas, fulfill reasonable physical constraints (see below). Our study is also of conceptual interest and it allows us to build the Rainich-like theory for the Einstein-Synge solutions. 

Synge EoS (4) implies that the thermodynamics of the fluid needs to be formulated in terms of the modified Bessel functions. Hence, it will not have a simple analytical expression. For this reason, different approximations have been considered in the literature in order to study the relativistic gas. 

On the one hand, some authors have used the limiting behavior of the modified Bessel functions in order to obtain an approximation of the Synge EoS for high and low temperatures \cite{Thorne} (see also \cite{Rezzolla} and references therein). 

On the other hand, Taub \cite{Taub} found that the pressure $p$, energy density $\rho$ and rest-mass density $n$ of a simple gas must satisfy a certain inequality in order to be consistent with kinetic theory, namely,
\be \label{Taub}
\hspace{-5mm} {\rm Ta} : \qquad \qquad  \qquad   \rho (\rho - 3p) \geq n^2 \, .  \qquad \qquad \qquad
\ee
Then, Mathews \cite{Mathews} used equality as the equation of state for a relativistic gas, and Mignone {\em et al.} \cite{Mignone, Mignone-2007} showed it to be a reasonable approximation to the Synge EoS. 

From now on we will refer to such an equation as the Taub-Mathews (TM) equation of state. Choi and Wiita \cite{Choi} used this approximation to construct an EoS for a multi-component relativistic gas suitable for numerical (special) relativistic hydrodynamics, which has been used recently to study the process of jet deceleration \cite{Perucho_2021}. Sokolov {\em et al.} \cite{Sokolov} also proposed an approximation to the Synge EoS, which does not fulfill the Taub constraint \cite{Mignone}.

The second goal of this paper is to analyze all these approximations and to introduce new ones. Our hydrodynamic approach (through the indicatrix function $\chi(\pi)$) enables us to study and compare the accuracy of the different approximations and to analyze whether they fulfill reasonable constraints for physical reality.

Pleba\'nski \cite{Plebanski} {\em energy conditions} are necessary algebraic conditions for physical reality and, in the perfect fluid case, they state $-\rho < p \leq \rho$. The determination of the spacetime regions where these hydrodynamic constraints hold is a basic task in analyzing a given perfect fluid solution. And a basic physical requirement imposed on the thermodynamic schemes is the positivity of the rest-mass density, of the temperature and of the specific internal energy, $\Theta > 0$, $\rho > n > 0$ ({\em positivity conditions}).

Moreover, in order to obtain a coherent theory of shock waves for the fundamental system of perfect fluid hydrodynamics, one must impose the relativistic compressibility conditions  \cite{Israel, Lichnero-1}:
\begin{eqnarray}
\hspace{-5mm} {\rm H}_1 : \qquad \qquad    (\tau'_p)_s < 0 \, , \qquad \quad (\tau''_p)_s > 0 \, , \ \qquad \\[2mm]
 \label{cc-1}
 %
\hspace{-10mm} {\rm H}_2 : \qquad \qquad \qquad \qquad   (\tau'_s)_p > 0 \, , \qquad \qquad \qquad \,
\label{cc-2}
\end{eqnarray}
where the function of state $\tau = \tau(p, s)$ is the {\em dynamic volume}, $\tau = \hat{h}/n$, $\hat{h} = h/c^2$ being the dimensionless enthalpy index. 

The above quoted physical constraints were inferred from macroscopic analysis. Moreover, we will also impose Taub's inequality Ta given in (\ref{Taub}) (which was inferred from the kinetic theory \cite{Taub}) on the equations of state that approximate that of a Synge gas.

In Sect. \ref{sec-flow-IG} we revisit the hydrodynamic approach to generic ideal gases. We analyze the ideal sonic condition for a perfect energy tensor $T$, and we give the ideal thermodynamic schemes associated with a $T$ that fulfills this hydrodynamic constraint. Moreover, the compressibility conditions H$_1$ and H$_2$, and Taub's inequality Ta are explicitly stated for the ideal gas case.

Sect. \ref{sec-synge} is devoted to achieving the first objective of this paper: the purely hydrodynamic labeling of Synge gas solutions. This result is applied  to analyze the behavior of a Synge gas at low and high temperatures and to establish the Rainich theory for the Einstein-Synge solutions.

In Sect. \ref{sec-approximations} we analyze equations of state that approximate the Synge EoS at low or at high temperatures, and use our hydrodynamic approach to determine acceptable approximations in the entire domain of applicability. In particular, we recover the Taub-Mathews EoS and we analyze its accuracy.

In Sect. \ref{sec-isentropic-evol} we consider the isentropic evolution of an ideal gas, and we apply this study to obtain the Friedmann equation for a TM ideal gas.

Finally, in Sect. \ref{sec-conclusions} we comment on the conceptual and practical interest of our results.

\section{Hydrodynamic flow of a generic ideal gas}
\label{sec-flow-IG}

If we are interested in a particular family of fluids defined by a specific equation of state we can analyze the fundamental system of the hydrodynamics for these fluids and: i) obtain a deductive criterion to detect whether a perfect energy tensor $T$ performs the evolution of a perfect fluid in this family, that is, obtain the specific hydrodynamic flow ({\em direct problem}), and ii) obtain all the perfect fluids in this family for which a $T$, fulfilling this criterion, gives a particular evolution ({\em inverse problem}).

In \cite{CFS-LTE} we have solved these problems for the paradigmatic family of ideal gases. Now we summarize these results, and we will apply them to study the Synge gas in Sec. \ref{sec-synge}, and to analyze the approximations to the Synge gas EoS in Sect. \ref{sec-approximations}.

	\subsection{Ideal hydrodynamic sonic condition}
\label{subsec-HSC-IG}

A generic ideal gas is characterized by the first equation of state given in (\ref{synge}), namely,
\begin{equation}
p = k n \Theta,   \qquad    k \equiv {k_B \over m}.  \label{gas-ideal}
\end{equation}
Then, the Duhem-Gibbs balance equation (\ref{re-termo}) implies that the specific energy $e = \rho/n$ is an effective function of the temperature, $e =e(\Theta)$, a function that characterizes each specific ideal gas. Thus, $\Theta = \Theta(e)$, and from (\ref{gas-ideal}) we obtain that the hydrodynamic variable $\pi=p/\rho$ is also a function of the specific energy $e$:
\begin{equation}
\pi = \pi(e) \equiv {k\Theta(e) \over e}  , \qquad  \pi \equiv \frac{p}{\rho}  \, .
\label{pi-e}
\end{equation}

For a generic (nonbarotropic) ideal gas we can take $(\rho,p)$ as coordinates in the thermodynamic plane \cite{CFS-LTE}. Moreover, we have $\pi'(e)\not=0$, and thus we can determine the inverse function $e=e(\pi)$. 
This will enable us to get a hydrodynamic characterization of the Synge gas, since with it, we can write all the thermodynamic quantities in terms of the hydrodynamic quantities $(\rho,p)$. We start by doing so for the speed of sound, which takes the expression \cite{CFS-LTE}:
\begin{equation}
c_s^2(\rho,p) = \pi + {1 \over \phi(\pi)}, \qquad  \phi(\pi) \equiv \frac{(\pi + 1) e'(\pi)}{\pi e(\pi)} . \label{vel-so}
\end{equation}
This expression shows that the hydrodynamic sonic condition (\ref{HSC}) imposes that the indicatrix function $\chi=c_s^2$ depends only on the hydrodynamic quantity $\pi$. This result solves the direct problem for the generic ideal gases \cite{CFS-LTE}:
\begin{itemize}
\item[]
The necessary and sufficient condition for a nonbarotropic and nonisoenergetic ($\dot{\rho} \not =0$) divergence-free energy tensor $T$ to represent the energy evolution of a generic ideal gas in local thermal equilibrium is that its hydrodynamic quantities $\{u,\rho,p\}$ satisfy $\chi = \chi(\pi) \not= \pi$, that is, they fulfill the {\em ideal sonic condition}:
\begin{subequations} \label{ISC}
\begin{eqnarray}
\hspace{-0mm} {\rm S^{\rm G}} :  \qquad \quad  	\dif \chi \wedge \dif \pi = 0 , \qquad \chi \not= \pi  \, , \label{dchi-dpi}  \quad  \qquad  \\[1mm]
	  \chi \equiv \frac{u(p)}{u(\rho)}  , \qquad \pi \equiv \frac{p}{\rho} \, . \quad  \qquad  \label{chi-pi} \end{eqnarray}
\end{subequations}
\end{itemize}
The above result states: (i) if  $\{u,\rho,p,n, s, \Theta\}$  is a solution of the fundamental system of the ideal gas hydrodynamics ${\cal F}_{\rm G} \equiv \{(\ref{nablaT}) (\ref{conservacio_massa}) (\ref{re-termo}) (\ref{gas-ideal})\}$, then $\{u,\rho,p\}$ is a solution of the ideal hydrodynamic flow system ${\cal{H}}_{\rm G} \equiv \{(\ref{nablaT}) (\ref{ISC})\}$, and conversely, (ii) if $\{u,\rho,p\}$ is a solution of the ideal hydrodynamic flow system ${\cal{H}}_{\rm G}$, then a solution $\{u,\rho,p,n, s, \Theta\}$ of the ideal fundamental system ${\cal F}_{\rm G}$ exists. The specific expression of these thermodynamic quantities in terms of the hydrodynamic ones is provided by the inverse problem, which was analyzed and solved in \cite{CFS-LTE}:
\begin{itemize}
\item[]
If a nonbarotropic and nonisoenergetic ($\dot{\rho} \not =0$) divergence-free energy tensor $T\equiv\{u,\rho,p\}$ satisfies (\ref{ISC}), then it represents the energy evolution of a generic ideal gas with specific energy $e$, temperature $\Theta$, rest-mass density $n$, and specific entropy $s$ given by:
\begin{subequations} \label{e-psi}
\begin{eqnarray}
e(\pi) = e_0 \exp\! \left[\int  \!\! \psi(\pi)d\pi \right] , \quad \label{e-pi} \\[0mm]
\psi(\pi) \equiv \frac{\pi}{(\pi+1)[\chi(\pi)-\pi]} \label{psi-pi}  , \quad
\end{eqnarray}
\end{subequations}
\be
\Theta(\pi) = {\pi \over k} e(\pi) , \quad n(\rho,p) = {\rho \over e(\pi)} ,  \qquad    \label{t-n-ideal}
\ee
\begin{subequations} \label{s-f-phi}
\begin{eqnarray}
s(\rho,p) = k \ln \frac{f(\pi)}{\rho} \, , \quad  \label{s-ideal} \\[0mm]
f(\pi) \equiv f_0 \exp\! \left[\int \! \! \phi(\pi) d\pi \right] ,    \quad  \label{f-pi} \\[0mm] 
\phi(\pi) \equiv \frac{1}{\chi(\pi)-\pi}  . \quad  \label{phi-pi} 
\end{eqnarray}
\end{subequations}
\end{itemize}
Note that the {\em ideal gas thermodynamic scheme} $\{(\ref{e-psi}) (\ref{t-n-ideal}) (\ref{s-f-phi})\}$ associated with a solution $T$ of the ideal flow system ${\cal{H}}_G$ depends on the real parameters $e_0$ and $f_0$. The first one, $e_0$, should be taken such that $e(0)= 1$ because $\rho=n$ at zero temperature. The second one, $f_0$, fixes the additive constant that determines the specific entropy for a given temperature.

It is worth remarking that the richness of thermodynamic schemes associated with any thermodynamic energy tensor $T$ by the inverse problem depends on two arbitrary functions of the specific entropy \cite{CFS-LTE}. Nevertheless, only the ideal thermodynamic scheme $\{(\ref{e-psi}) (\ref{t-n-ideal}) (\ref{s-f-phi})\}$ is compatible with the equation of state (\ref{gas-ideal}) of a generic ideal gas.

\subsection{Constraints for physical reality}

For an ideal gas, the equation of state (\ref{gas-ideal}) and the positivity of the temperature and of the rest-mass density imply a positive thermodynamic pressure, $p > 0$. Consequently, the energy conditions become:
\begin{equation} \label{e-c-G}
\hspace{4mm} {\rm E}^{\rm G} : \qquad     \rho > 0 \, , \qquad  0 < \pi \leq 1 \, , \quad  \pi = p/\rho \, .
\end{equation}

On the other hand, in \cite{CFS-CC} we have proved that the compressibility conditions H$_1$ constrain the hydrodynamic evolution of the fluid, and for an indicatrix function of the form $\chi = \chi(\pi)$, they can be written as:
\begin{equation}  \label{H1G}
\hspace{4mm} {\rm H}_1^{\rm G} : \quad 
\begin{array}{c}      
0 < \chi < 1  ,  \qquad \quad  \\[2mm] 
\zeta \equiv (1+\pi)(\chi-\pi) \chi'  + 2 \chi(1-\chi) > 0  .  \
\end{array}
\end{equation}
In addition, the remaining compressibility condition H$_2$ constrains the associated thermodynamic schemes $\{s,n,\Theta\}$ \cite{CFS-CC}. However, for the ideal thermodynamic scheme \{(\ref{e-psi})(\ref{t-n-ideal})(\ref{s-f-phi})\} condition H$_2$ can also be expressed in terms of the indicatrix function $\chi(\pi)$ as \cite{CFS-CC}:
\begin{equation} \label{H2G}
{\rm H}^{\rm G}_2: \qquad \qquad \xi \equiv (2\pi + 1)\chi - \pi > 0 \, . \qquad \qquad 
\end{equation}

Finally, Taub's inequality Ta given in (\ref{Taub}) can be written for the ideal gas case as:
\begin{equation} \label{TaubG}
{\rm Ta}^{\rm \! G}: \quad \quad \ \qquad  \eta \equiv e^2(\pi)(1 - 3\pi)  \geq 1 \, . \qquad \quad
\end{equation}

In order to analyze when a specific function of state $\chi(\pi)$ corresponds to a physically realistic ideal gas, we must analyze its behavior in the domain $]0, 1]$ where the energy conditions  E$^G$ hold. This behavior should be compatible with the compressibility conditions H$_1^{\rm G}$ and H$_2^{\rm G}$. Moreover, the specific energy $e(\pi)$ given by (\ref{e-psi}) should fulfill Taub's inequality Ta$^{\rm \! G}$.

\section{Hydrodynamic flow of a Synge relativistic gas}
\label{sec-synge}

A Synge gas is characterized by two equations: the EoS of a generic ideal gas (\ref{gas-ideal}) and the second one in Eq. (\ref{synge}), namely, 
\begin{equation} \label{Synge-2}
\rho + p = n \, h(z)  ,  \quad z \equiv {1 \over k \Theta},   \qquad     h(z)\equiv {K_3(z) \over K_2(z)} ,  
\end{equation}
$K_n$ being the modified Bessel functions of the second kind. It is worth remarking that the relation between the pair of thermodynamic quantities $(z,h)$ given in (\ref{Synge-2}) is not the only way to characterize the Synge gas. This equation of state leads to relations between other pairs of quantities which also characterize the Synge gas. Now, the objective is to find one of these equations of state that only involves hydrodynamic quantities.

We know (see Sec. \ref{sec-flow-IG}) that, as a consequence of the ideal gas EoS (\ref{gas-ideal}), the speed of sound is a function of the hydrodynamic quantity $\pi =p/\rho$, $c_s^2 = \chi(\pi)$. Now we show that the Synge equation of state (\ref{Synge-2}) determines $\chi(\pi)$ by imposing a specific first-order differential equation.

	\subsection{Hydrodynamic characterization of the Synge gas}
\label{subsec-chi-synge}

Based on the properties of the Bessel functions, the Synge equation (\ref{Synge-2}) can be written as  
\begin{equation}
h(z) = {2 \over z} - {K_2'(z) \over K_2(z)}  .  \label{h-K2}
\end{equation}
The function $K_2(z)$ is a solution of the Bessel equation,
\begin{equation}
z^2 K_2''(z) + z K_2'(z) - (4+z^2)K_2(z) = 0 ,    \label{bessel}
\end{equation}
the only one fulfilling the boundary condition $K_2(\infty)=0$. Then, the function $h=h(z)$ may be characterized by the following differential equation equivalent to (\ref{bessel}):
\begin{equation}
z[h'(z)-h^2+1] + 5 h  = 0 ,       \label{bessel-h}
\end{equation}
together with the boundary condition $h(\infty)=1$. Moreover, this condition ensures that $h(z)$ is greater than $1$ in all the domain, as required by the definition (\ref{re-termo}) of the relativistic specific enthalpy and the positivity conditions. Using Eqs. (\ref{gas-ideal}) and (\ref{t-n-ideal}) we can obtain the function of state that relates the quantities $(z, \pi)$: 
\begin{equation} \label{pi-z}
\pi = \frac{1}{z \, e} = \frac{1}{z h(z) - 1} \equiv \pi (z) \, .
\end{equation}
Eqs. (\ref{bessel-h}) and (\ref{pi-z}) give us the differential equation that, together with the conditions $\pi(\infty)= 0^+$ and $\pi'(\infty)= 0$, characterize the function $\pi= \pi(z)$:
\begin{equation} \label{bessel-pi}
z \, \pi'(z)-(3 + z^2)\pi^2 - 2\pi + 1 = 0 \, .
\end{equation}
In this case, the boundary conditions ensure that the function $\pi=\pi(z)$ takes values in the interval $]0, 1/3]$, a fact that is compatible with the energy conditions E$^{\rm G}$. 
On the other hand, by deriving the first equality in (\ref{pi-z}) with respect to $z$ and using (\ref{e-psi}) and (\ref{pi-z}) to eliminate $e(\pi)$ and $e'(\pi)$, we have
\begin{equation} \label{z pi-prima}
z \, \pi'(z) = \frac{\pi (\pi + 1)[\pi - \chi(\pi)]}{\pi [\chi(\pi) - 1] + \chi(\pi)} \, ,
\end{equation}
which can be substituted in (\ref{bessel-pi}) to give
\begin{equation} \label{z2-pi}
z^2 = \frac{1}{\pi^2}-\frac{3}{\pi}+\frac{\pi}{\pi [\chi(\pi) - 1] + \chi(\pi)} - 3 \, .
\end{equation}
Finally, deriving (\ref{z2-pi}) with respect to $z$ and using (\ref{z pi-prima}) we obtain that the indicatrix function of a Synge gas is the solution to the first-order differential equation:
\begin{subequations} \label{chi-prima}
\begin{equation} \label{chi-prima-eq}
\chi'(\pi) = {\cal S}(\chi, \pi) \equiv \frac{\alpha  \chi^3 + \beta  \chi^2 + \gamma  \chi + \delta}{\pi^3 (\pi + 1) (\pi - \chi)} \, ,
\end{equation}
\begin{eqnarray}
\alpha & = & \alpha(\pi) \equiv 3(1 + 4\pi + 5\pi^2 + 2\pi^3) \, , \label{subeq-alpha} \\[2mm]
\beta & = & \beta(\pi) \equiv -\pi(11 + 29\pi + 17\pi^2) \, , \label{subeq-beta} \\[2mm]
\gamma & = & \gamma(\pi) \equiv \pi^2(13 + 17\pi) \, , \label{subeq-gamma} \\[2mm]
\delta & = & \delta(\pi) \equiv -5\pi^3 \, , \label{subeq-delta}
\end{eqnarray}
\end{subequations}
with the boundary condition $\chi'(0) = 5/3$. Then, $\chi(0)=0$, and the solution $\chi=\chi(\pi)$ takes values in the interval $[0,1/3]$, which is compatible with the first of the compressibility conditions H$^{\rm G}$. 
%

This statement certainly gives us a hydrodynamic characterization of the Synge gas because it only uses conditions on the hydrodynamic quantities $\{u, \, \rho, \, p\}$. But it is not a deductive characterization because the existence of a functional dependence between the variables $\chi$ and $\pi$ does not imply that the expression of the function $\chi(\pi)$ is known, which is a necessary requirement in order to impose condition (\ref{chi-prima}). Nevertheless, a deductive characterization easily follows:
\begin{itemize}
\item[]
The necessary and sufficient condition for a non isoenergetic ($\dot{\rho} \not =0$) divergence-free energy tensor $T$ to represent the energy evolution of a Synge gas is that its hydrodynamic quantities $\{u,\rho,p\}$ fulfill the {\em Synge sonic condition}:
\begin{equation} \label{SyngeSC}
\hspace{-0mm} {\rm S^{\rm S}} :  \quad \qquad  \ \  	\dif \chi =  {\cal S}(\chi, \pi) \dif \pi  \, , \quad  
\end{equation}
where ${\cal S}(\chi, \pi)$ is given in \rm(\ref{chi-prima}), and $S(0,0) = 5/3$.
\end{itemize}
The above result states: (i) if  $\{u,\rho,p,n, s, \Theta\}$  is a solution of the fundamental system of the Synge gas hydrodynamics ${\cal F}_{\rm S} \equiv \{(\ref{nablaT}) (\ref{conservacio_massa}) (\ref{re-termo}) (\ref{gas-ideal}) (\ref{Synge-2})\}$, then $\{u,\rho,p\}$ is a solution of the Synge hydrodynamic flow system ${\cal{H}}_{\rm S} \equiv \{(\ref{nablaT}) (\ref{SyngeSC})\}$, and conversely, (ii) if $\{u,\rho,p\}$ is a solution of the ideal hydrodynamic flow system ${\cal{H}}_{\rm S}$, then a solution $\{u,\rho,p,n, s, \Theta\}$ of the Synge fundamental system ${\cal F}_{\rm S}$ exists.

A specific ideal gas, and in particular a Synge gas, is defined by a specific indicatrix function $\chi(\pi)$, but also by the function $e=e(\pi)$. The former one is an explicit hydrodynamic quantity, $\chi = u(p)/u(\rho)$, and for that reason it is the right one to give the above hydrodynamic characterization. Nevertheless, it can also be conceptually interesting to characterize a Synge gas in terms of the pair $(\pi, e)$. 
From (\ref{t-n-ideal}) and  (\ref{Synge-2}) we obtain:
\be
e(z) = h(z) - \frac{1}{z} \, .
\ee
Using this equation to write Eq. (\ref{bessel-h}) for $e=e(z)$, we get:
\be \label{e'-1}
z^2 e'(z) + z^2[1-e^2(z)] + 3[1+ze(z)] = 0  \, ,
\ee
while, from (\ref{pi-z}) we obtain:
\be \label{e'-2}
e'(\pi) = - \frac{\pi}{e} [1+ \pi e^2 z'(e)] = - \pi \frac{\pi e^2 + e'(z)}{e e'(z)} \, .
\ee
Then, from Eqs. (\ref{e'-1}) and (\ref{e'-2}) we have that the function $e=e(\pi)$ of a Synge gas fulfills the boundary condition $e'(0) =3/2$ and the first-order differential equation:
\begin{equation}
e'(\pi )= \frac{e(\pi)[1+e^2(\pi)(3 \pi^2 \!+ \! 3 \pi \! - \!1)]}{ \pi [e^2(\pi)(1\!-\!2\pi\!-\!3\pi^2)-1]} \, .   \label{e'-pi}
\end{equation}

Another function of state that characterizes an ideal gas is the so-called (generalized) adiabatic index $\Gamma \equiv (\partial \ln n/ \partial \ln p)_s$. For a generic ideal gas it is related to the indicatrix function $\chi(\pi)$ by the expression \cite{Rezzolla, Krautter}:
\be \label{Gamma}
\Gamma(\pi) = \frac{1+\pi}{\pi}\chi(\pi) \, .
\ee
Note that this expression shows that $\Gamma$ can be obtained from the hydrodynamic quantities $(u, \rho, p)$, and thus it can be used to identify a specific ideal gas. Equations (\ref{chi-prima}) and (\ref{Gamma}) imply that the $\Gamma =\Gamma(\pi)$ of a Synge gas is characterized by the first-order differential equation:
\be
\Gamma'(\pi) = \frac{(\Gamma-1)^2[5(\pi+1) - 3 \Gamma (2 \pi +1)]}{\pi^2(\Gamma-1-\pi)} \, ,
\ee
and the boundary condition $\Gamma'(0) = -5/3$.

	\subsection{Behavior of the Synge gas at low and high temperatures}
\label{subsec-approaches}

In the previous subsection we have characterized the indicatrix function  $\chi(\pi)$ of a Synge gas through the differential equation (\ref{chi-prima}), which cannot be solved analytically. Now, in this subsection, we will use it in order to study the behavior of $\chi(\pi)$ in the limiting cases at low and high temperatures. 

On the one hand, the limit $\Theta=0$ corresponds to $p=0$, and then $\pi=0$. On the other hand, the limit at high temperature corresponds to $z=0$, and from expressions (\ref{h-K2}) and (\ref{pi-z}) of $h(z)$ and $\pi(z)$, to $\pi = 1/3$. Thus, the indicatrix function of a Synge gas $\chi(\pi)$ takes values in the domain $[0, \, 1/3]$. 

Note that Eq. (\ref{chi-prima}) determines the value of $\chi(\pi)$ at both ends of the interval. Indeed, as commented in subsection above, we have that 
\begin{equation} \label{chi-en-zero}
\chi(0) = 0 \, , \quad \qquad \chi'(0) = 5/3 \, .
\end{equation}
Then, these values and the successive derivatives of Eq. (\ref{chi-prima}) determine the derivatives of $\chi(\pi)$ at $\pi=0$. This fact enables us to know the behavior of the indicatrix function $\chi(\pi)$ of a Synge gas at low temperatures by writing its Taylor expansion around $\pi = 0$. For instance, the second, third and fourth derivatives take the values:
\begin{equation} \label{derivades-chi-zero}
\chi''(0) = -\frac{20}{3} \, , \quad \chi'''(0) = 45 \, ,\quad \chi^{iv}(0) = 460 \, .
\end{equation}

Based on the behavior of the Bessel functions close to zero, it is also possible to get that
\begin{equation} \label{derivades-chi-1/3}
\chi(1/3) = 1/3 \, , \qquad \chi'(1/3) = 1/2 \, ,
\end{equation}
which is compatible with Eq. (\ref{chi-prima}). However, as can be seen from these values together with the derivative of Eq. (\ref{chi-prima}), the second derivative of $\chi(\pi)$ is not defined at $\pi = 1/3$. This is a consequence of the fact that the function $K_2(z)$ has a singular point at $z=0$. Thus, we can only know the behavior of the indicatrix function $\chi(\pi)$ for a Synge gas around $\pi = 1/3$ (at high temperatures) up to first order in Taylor expansion.

\subsection{Rainich-like theory for the Einstein-Synge solutions}
\label{subsec-Rainich} 

An Einstein-Synge solution is a solution of the Einstein field equations, $G(g) = \kappa T$, where $T$ is a perfect energy tensor that models the energy evolution of a Synge gas. The Einstein-Synge equations involve the metric tensor $g$ and the thermodynamic quantities $\{u,\rho,p,n, s, \Theta\}$ that are constrained by the fundamental system of the Synge gas hydrodynamics ${\cal F}_{\rm S} \equiv \{(\ref{nablaT}) (\ref{conservacio_massa}) (\ref{re-termo}) (\ref{gas-ideal}) (\ref{Synge-2})\}$.  

Note that, as a consequence of the hydrodynamic characterization obtained in the subsection above, the Einstein-Synge equations are equivalent to a system that only involves the variables $\{g, u, \rho, p\}$: the field equations $G(g) = \kappa T$, $T \equiv \{u, \rho, p\}$ being a solution of the Synge hydrodynamic flow system ${\cal{H}}_{\rm S} \equiv \{(\ref{nablaT}) (\ref{ISC})(\ref{SyngeSC})\}$. Then, a question naturally arises: is there a system of equations involving only the metric tensor $g$ that characterizes the Einstein-Synge solutions? Answering that question and obtaining those equations means giving the Rainich-like theory for the Einstein-Synge solutions.

The first theory that characterizes the nonvacuum solutions of the Einstein field equations corresponding to a specific energy content was formulated by Rainich \cite{Rainich}. He gave the necessary and sufficient conditions for a metric tensor $g$ to be an Einstein-Maxwell solution for a regular electromagnetic field. 
 
In \cite{Coll-Ferrando-termo} we developed a similar theory for the thermodynamic perfect fluid solutions of the Einstein equations, and the hydrodynamic sonic condition plays an important role in that study.  Later,  we gave in \cite{CFS-CC} the necessary and sufficient conditions for a perfect fluid solution to describe a generic ideal gas that fulfills the compressibility conditions H$_1^G$ and H$_2^G$. 

If we want the fluid to be a Synge gas, Eq. (\ref{SyngeSC}) must be added to those conditions. Nevertheless, now H$_1^G$ and H$_2^G$ identically hold, and they can be removed in the characterization statement. Thus, if we take into account theorems 9 and 10 in reference \cite{CFS-CC} we can obtain the Rainich-like theory for the Einstein-Synge solutions. 
Below, we present the explicit expressions of this invariant labeling.

\noindent
{\bf Ricci concomitants}: consider the following scalar and tensor functions of the Ricci tensor $R$ and its derivatives:
\begin{eqnarray}
t \equiv \tr R  , \qquad  \qquad  N \equiv R - \frac14 t \, g  ,\label{fluper-definitions-1} \qquad \\[-1mm]   \displaystyle q \equiv - 2 \sqrt[3]{\frac{\tr N^3}{3}} , \qquad Q \equiv N - \frac14 q \, g , \label{fluper-definitions-2} \qquad  \\[-1mm]
\label{fluper-hydro}
 \rho = \frac14 (3 q + t)   , \quad \qquad p = \frac14 (q - t)  , \qquad \\[0mm]
\pi  \equiv \frac{p}{\rho}   , \quad \chi  \equiv \frac{Q( \dif p , \dif \rho)}{Q( \dif \rho , \dif \rho)}   , \quad  {\cal S} \equiv {\cal S}(\chi, \pi), \label{Rainich-Synge}
\end{eqnarray}
%
%
where ${\cal S}(\chi, \pi)$ is given in \rm(\ref{chi-prima}). \\[2mm]
{\bf Characterization theorem 1}: a metric is an Einstein-Synge solution in nonisoenergetic evolution if, and only if, the Ricci tensor $R$ satisfies the invariant conditions:
\begin{eqnarray} \label{fluper-conditions-ideal}
Q^2 + q Q = 0  , \quad  Q(y,y) > 0   ,  \quad -t < q \leq t   ,  \qquad \ \\[2mm]
Q(\dif \rho) \not=0, \quad   \dif \chi =  {\cal S}(\chi, \pi) \dif \pi   , \quad {\cal S}(0,0)= 5/3  , \qquad \  \label{Si(x)}
\end{eqnarray}
where $y$ is any timelike vector, and where $Q$, $N$, $q$, $t$, $\rho$, $p$, $\pi$, $\chi$ and ${\cal S}$ are given in (\ref{fluper-definitions-1}), (\ref{fluper-definitions-2}), (\ref{fluper-hydro}) and (\ref{Rainich-Synge}).\\[1mm]
{\bf Characterization theorem 2}: a metric is an Einstein-Synge solution in isoenergetic evolution if, and only if, the Ricci tensor $R$ satisfies the invariant conditions given in (\ref{fluper-conditions-ideal}) and:
\begin{equation} \label{fluper-conditions-isoenergetic}
Q(\dif \rho) = 0  , \qquad  Q(\dif p) = 0   ,
\end{equation}
where $y$ is any time-like vector, and $Q$, $N$, $q$, $t$, $\rho$ and $p$ are given in (\ref{fluper-definitions-1}), (\ref{fluper-definitions-2})  and (\ref{fluper-hydro}). 
%

\section{Approximations to the Synge fluid}
\label{sec-approximations}
The aim of this section is to take advantage of the results of Sec. \ref{sec-synge} in order to try to find some analytical expressions for $\chi(\pi)$ approximating that of a Synge gas, namely, the solution to the differential equation (\ref{chi-prima}).

	\subsection{Classical ideal gas approximation}
\label{subsec-CIG-approx}

Classical ideal gases are the ideal gases (EoS (\ref{gas-ideal})) fulfilling the $\gamma$\textit{-law} equation of state $p = (\gamma - 1)n\epsilon$, where $\epsilon = e(\Theta) - 1$ is the specific internal energy and $\gamma$ the adiabatic index. They are usually considered as a good approximation of an ideal gas at low temperatures, and they can equivalently be characterized as those with an indicatrix function of the form \cite{CFS-CIG}:
\be
\chi_c(\pi) = \frac{\gamma \pi}{\pi +1} \, . \label{chi-CIG} 
\ee

The adiabatic index $\gamma = 5/3$ corresponds to a mono\-atomic gas. In this case $\chi_c(0)=0$, $\chi_c'(0) = 5/3$, and thus it approaches a Synge gas at first order. For the sake of completeness, we analyze the constraints for physical reality for any $\gamma$.

In \cite{CFS-CIG} we have studied the macroscopic compressibility conditions in this case and we have obtained: 
\begin{itemize}
\item[]
For a classical ideal gas with $\Theta > 0$, $\rho > n > 0$, the macroscopic constraints for physical reality E$^G$, H$^G_1$ and H$^G_2$ are satisfied for values of $\pi$ in a nonempty subinterval of $[0,1]$ if, and only if, adiabatic index fulfills $\gamma >1$. In fact, they hold in the interval:
\begin{equation} \label{cc-CIG}
 \begin{cases}  0 < \pi <   {\pi}_m \equiv  \gamma -1 \, , \quad {\rm if} \ \ 1< \gamma \leq 2 \, ,  \cr
\displaystyle 0 < \pi <   \tilde{\pi}_m \equiv  \frac{1}{\gamma -1}
 , \quad {\rm if} \ \  \gamma \geq 2 \, . 
\end{cases}
\end{equation}
\end{itemize}

Usually (see, for example \cite{Anile}) the adiabatic index $\gamma$ is considered to be constrained by 
$1 < \gamma  \leq 2$. The statement above shows that, under reasonable macroscopic physical requirements, the upper limit for $\gamma$ can be relaxed. Note that the interval defined in (\ref{cc-CIG}) contains the domain $[0, 1/3]$ if $4/3 \leq \gamma \leq 4$. 

Nevertheless, it is known \cite{Taub, Rezzolla} that the relativistic kinetic theory imposes stronger restrictions. Indeed, Taub's inequality Ta$^{\rm \!G}$ given in (\ref{TaubG}) now takes the expression:
\begin{equation} \label{TaubC}
{\rm Ta}^{\rm \! C}: \qquad \quad  \   \pi < \hat{\pi}_m \equiv (\gamma-1)(5-3\gamma) \, . \qquad \quad   
\end{equation}
Note that if this constraint holds for some $\pi>0$, then necessarily $1 <\gamma <5/3$. Moreover, the maximum value for $\hat{\pi}_m$ is $1/3$, and it is reached when $\gamma = 4/3$.

Consequently, the classical monoatomic gas ($\gamma=5/3$) approximates a Synge gas at first order at low temperatures, and it fulfills the macroscopic constraint for physical reality H$^G_1$ and H$^G_2$ in the interval $[0, \, 2/3[$. Nevertheless, it does not fulfill Taub's inequality at any point.

	\subsection{An approximation at high temperature}
\label{subsec-high-approx}

The ultrarelativistic limit of a monoatomic gas is usually obtained from the Synge EoS by making the limit of the Bessel function $K_2(z)$ at $z = 0$ (see, for example, \cite{Rezzolla}). The thermodynamic quantities fulfill the relations:
\begin{equation}
\rho = a \Theta^4  , \ \quad  p = \frac13 a \Theta^4  , \ \quad  S=n s_1 = \frac43 a \Theta^3   ,
\label{s-t-e-radiacio}
\end{equation}
where $a$ and $s_1$ are constant. 

On the other hand, the above thermodynamic scheme can be also obtained by considering that the fluid fulfills the barotropic equation of state $\rho = 3p$, and using the usual macroscopic thermodynamic reasoning \cite{CFS-LTE}. Now, we will see that this ultrarelativistic limit can also be obtained from our hydrodynamic approach, that is, as an approximation from the Synge indicatrix function $\chi(\pi)$. 

Let us consider the zero-order approximation at $\pi=1/3$, that is, $c_s^2 = \chi(\pi) =1/3$. This indicatrix function corresponds to a specific nonbarotropic ideal gas that, applying the expressions (\ref{e-psi}) and (\ref{s-f-phi}) of the ideal inverse problem, fulfills the following equations of state:
\begin{subequations} \label{e-s-high}
\begin{eqnarray}
e(\pi) = e_0 [(\pi +1) ^3(1-3\pi)]^{-1/4}  , \quad \label{e-pi-high} \\[2mm]
s(\rho,p) = s_0 - k \ln(\rho-3p) \, .  \label{s-high}   \quad
\end{eqnarray}
\end{subequations}
If we impose that the ideal gas defined by Eqs. (\ref{e-s-high}) has an isentropic evolution, $s(\rho,p)= s_1$ = constant, then it fulfills the following barotropic (evolution) relation:
\be
p = \frac13(\rho - \kappa) \, , \qquad \kappa = {\rm constant} \, .
\ee
Then, the behavior of this model when $\kappa \rightarrow 0$ is that of the scheme given in Eq. (\ref{s-t-e-radiacio}). Thus, this usual ultrarelativistic limit to a Synge gas can be obtained as the limit of a one-parametric family of isentropic evolutions of a nonbarotropic ideal gas. 

A similar reasoning would allow us to interpret the so-called relativistic $\gamma$-law models, $p=(\gamma -1) \rho$, as the limit of a family of isentropic evolutions of the nonbarotropic ideal gas defined by the indicatrix function $c_s^2 = \chi(\pi) =\gamma -1$.

\subsection{Taub-Mathews approximation}
\label{subsec-TM-approx}

From now on, we consider equations of state that are Pad\'e-like approximants at $\pi=0$ and $\pi=1/3$, which approximate the indicatrix function $\chi(\pi)$ of a Synge EoS in the entire domain $[0,1/3]$. Let us start by considering for $\chi(\pi)$ a quotient P2/P1, where Pn denotes a polynomial of degree n:
\begin{equation} \label{ansatz-TM}
\chi(\pi) = \frac{\pi^2 + c_1\pi + c_2}{c_3\pi + c_4} \, ,
\end{equation}
$c_i$ being arbitrary constants with, at least, $c_3 \neq 0$. 

Thus, we can impose up to four conditions on the indicatrix function (\ref{ansatz-TM}). Then, we can use the results of the previous subsection in order to make sure that our approximated indicatrix function behaves as that of a Synge gas up to first order, at both low and high temperatures. In other words, we can make (\ref{ansatz-TM}) verify $\chi(0) = 0$, $\chi(1/3) = 1/3$, $\chi'(0) = 5/3$ and $\chi'(1/3) = 1/2$. A straightforward calculation shows that by doing so, we get:
\begin{equation} \label{chi-TM}
\chi(\pi) = \frac{\pi(5 - 3\pi)}{3(1 + \pi)} \, .
\end{equation}

In order to check whether the fluid with the above indicatrix function verifies the compressibility conditions $\rm{H}_1^{\rm{G}}$ and $\rm{H}_2^{\rm{G}}$ we simply have to substitute (\ref{chi-TM}) in (\ref{H1G}) and (\ref{H2G}), respectively. By doing so, we get:
\begin{eqnarray} \label{H1G-TM}
\zeta = \frac{8\pi(5 - 10\pi + 9\pi^2)}{9(1 + \pi)^2} > 0 \, , \\
 \label{H2G-TM}
\xi = \frac{2\pi (1 + 2\pi - 3\pi^2)}{3 (1 + \pi)} > 0 \, ,
\end{eqnarray}
if $\pi \in ]0,1/3[$. So, both compressibility conditions are fulfilled.

Now, we can determine the ideal thermodynamic scheme by using (\ref{e-psi}) and (\ref{s-f-phi}), and we obtain:
\begin{equation} \label{e-f-TM}
e(\pi) = \frac{1}{\sqrt{1 - 3\pi}} \, , \qquad f(\pi) = f_0 \frac{\pi^{\frac32}}{(1 - 3\pi)^2} \, ,
\end{equation}
where we have set $e_0 = 1$ so that $\epsilon(0) = e(0) - 1 = 0$. Then, we obtain the following equation of state: 
\begin{equation} \label{TM-1}
e^2(\pi)(1 - 3\pi)  = 1 \, . \qquad \quad
\end{equation}
Note that this equation of state verifies Taub's inequality given in (\ref{TaubG}). In fact, it fulfills equality. 

On the other hand, by inverting $e = e(\pi)$ in (\ref{e-f-TM}) and using the definition of $\pi$ and (\ref{t-n-ideal}), we get that (\ref{TM-1}) becomes
\begin{equation} \label{TM-2}
p = \frac13 n \left(e - \frac1e \right) \, ,
\end{equation}
which is the equation of state proposed by Mathews \cite{Mathews} (TM EoS). Moreover, using (\ref{gas-ideal}) and (\ref{t-n-ideal}), it can be rewritten as
\begin{equation} \label{TM-EoS-ht}
(h - k \Theta)(h - 4k \Theta) = 1 \, ,
\end{equation}
which was proposed and analyzed by Mignone {\em et al.} \cite{Mignone, Mignone-2007}, who showed it to be a reasonable approximation to the Synge equation of state.

\begin{figure*}
\center
\includegraphics[width=0.7\textwidth]{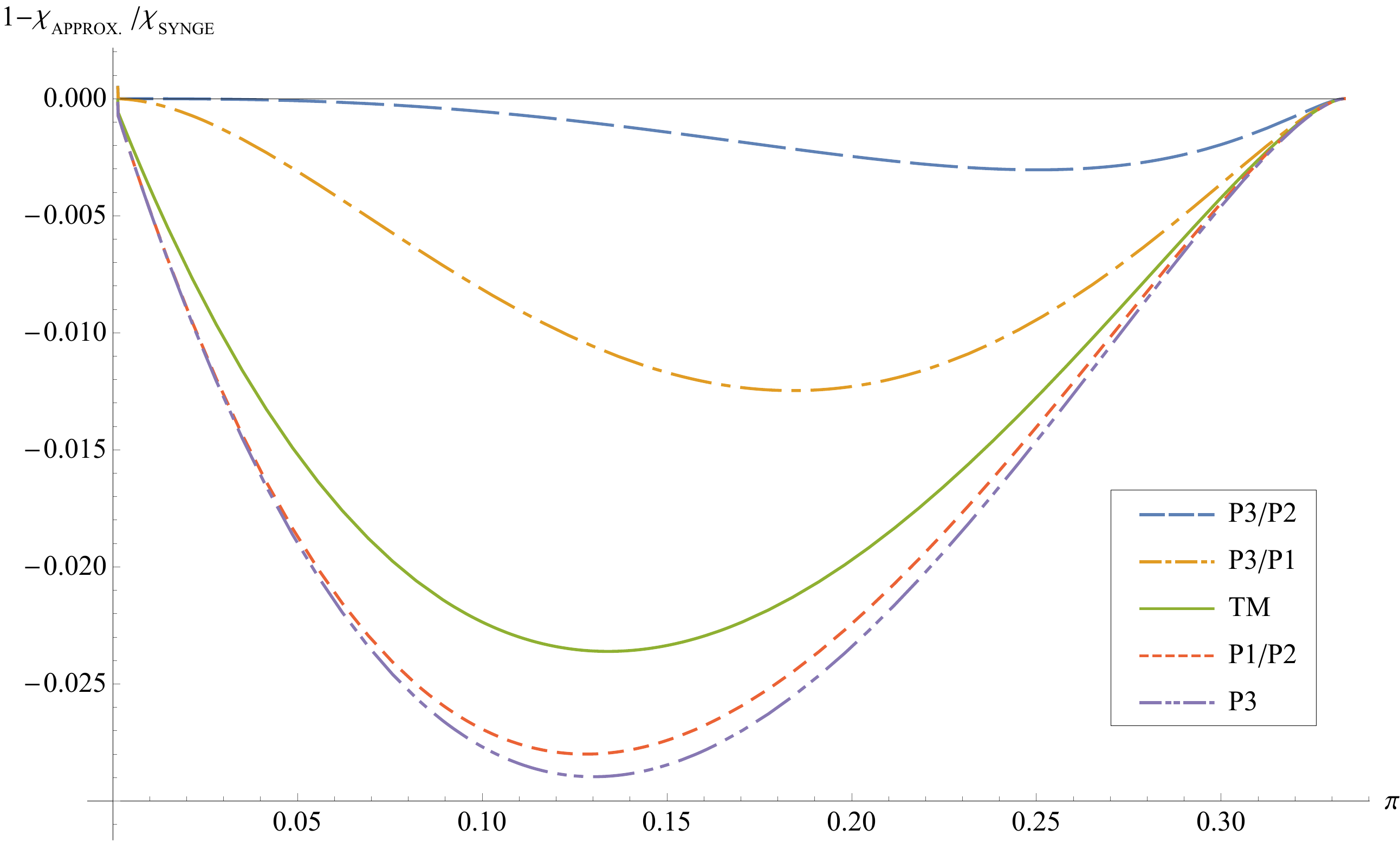}
\caption{This plot shows the relative error with respect to the Synge EoS of the TM EoS and of the other proposed EoS. The P3/P2 and P3/P1 EoS are more accurate than the TM EoS, while P3 and P1/P2 EoS are not.}
\label{fig-1}
\end{figure*}

If we compare the indicatrix function of the TM EoS with that of a Synge gas, we conclude that the relative error is less than 2.36$\%$ in the entire domain $[0, 1/3]$ (see Fig. \ref{fig-1}).

%

	\subsection{Generalized Taub-Mathews approximation}
\label{subsec-generalized-TM-approx}

We can try to generalize the TM EoS to an arbitrary $\gamma$ in order to obtain an analytical model for a relativistic polyatomic gas. We start from an expression of the form (\ref{ansatz-TM}) and we impose $\chi'(0) = \gamma$ instead of $\chi'(0) = \frac53$. In that case, we get the following indicatrix function (TM$\gamma$):
\begin{equation} \label{chi-TMgamma}
\chi(\pi) = \frac{\pi[\gamma + 3\pi (\gamma - 2)]}{1+3\pi (2\gamma - 3)} \, ,
\end{equation}
which indeed reduces to (\ref{chi-TM}) for $\gamma = \frac53$. 

Now, we can easily check that $\zeta>0$ and $\xi >0$, and thus both compressibility conditions (\ref{H1G}) and (\ref{H2G}) are fulfilled for any $\gamma \geqslant 1$.

Moreover, following the same procedure as before, we have that, in this case:
\begin{equation} \label{e-f-TMgamma}
e(\pi) = \frac{(1 + \pi)^{\frac{5 - 3\gamma}{2(\gamma - 1)}}}{\sqrt{1 - 3\pi}} \, , \quad f(\pi) = f_0 \frac{\pi^{\frac{1}{\gamma - 1}}}{(1 - 3\pi)^2} \, ,
\end{equation}
Then, we can obtain: 
\begin{equation}
\eta \equiv e^2(\pi)(1 - 3\pi)  = (1 + \pi)^{\frac{5 - 3\gamma}{\gamma - 1}} \, ,
\end{equation}
which implies that Taub's inequality (\ref{TaubG}) is only verified if $1 < \gamma \leqslant \frac53$. These results and all those that will be obtained in this section are summarized in Table \ref{table-1}.

The first expression in (\ref{e-f-TMgamma}) can be written as:
\begin{equation} \label{TMgamma-EOS-pe}
(n e + p)^{\frac{3\gamma - 5}{\gamma - 1}}(n e - 3p) = e^{\frac{2(\gamma - 2)}{\gamma-1}} n^{\frac{2(2\gamma - 3)}{\gamma-1}} \, ,
\end{equation}
or, equivalently,
\begin{equation} \label{TMgamma-EOS-ht}
h^{\frac{3\gamma - 5}{\gamma - 1}}(h - k \Theta)^{\frac{2(2 - \gamma)}{\gamma - 1}}(h - 4k \Theta) = 1 \, .
\end{equation}

In \cite{Sokolov} Sokolov proposed the simplified equation of state:
\begin{equation}
h (h - 4k \Theta) = 1 \, .
\end{equation}
It can easily be seen that this is a particular case of the TM$\gamma$ equation of state with $\gamma = 2$. Therefore, it neither reproduces the correct behavior of the Synge EoS for low temperatures nor fulfills Taub's condition. Nevertheless, we consider it in Table \ref{table-1} for the sake of completeness.


\begin{table*}
\caption{This table summarizes the physical behavior of the equations of state that approximate the Synge EoS. The two first columns show the indicatrix function $\chi(\pi)$ and the specific energy $e(\pi)$, respectively. In the third and forth columns, a $\checkmark$ indicates that the corresponding EoS fulfills the macroscopic compressibility conditions \rm{H}$_1^{\rm{G}}$, \rm{H}$_2^{\rm{G}}$ or the Taub constraint Ta$^G$ in the full interval $]0, 1/3[$. In the two first rows, the EoS depends on the adiabatic index $\gamma$; then, we indicate the values of this parameter for which these conditions hold.}
\vspace{2mm}
\label{table-1}
\begin{tabular}{cccccccl}
 \hline \\[-3.8mm] \hline \\[-2.5mm]
  EoS$_{\rm APPROX}$ & $\chi(\pi)$ & $e(\pi)$ & $\qquad \rm{H}_1^{\rm{G}}, $  $\ \rm{H}_2^{\rm{G}} \qquad$ & $\qquad \rm{Ta^G} \qquad$ &
  \hspace{-10mm}  
  \\[1mm]
 \hline \\[-2mm]
   \  CIG & $\frac{\gamma \pi}{1 + \pi}$ & $\frac{\gamma - 1}{\gamma - 1 - \pi}$  &  $4/3 < \gamma < 4$ & $\gamma = \frac43$ &
\\[1.5mm]  
   \  TM$\gamma$ & $\frac{\pi[\gamma + 3\pi (\gamma - 2)]}{1+3\pi (2\gamma - 3)}$ & $\frac{(1 + \pi)^{\frac{5 - 3\gamma}{2(\gamma - 1)}}}{\sqrt{1 - 3\pi}}$  &    $\gamma > 1$ & $1 < \gamma \leqslant \frac53$ &
\\[2mm] 
   \  TM & $\frac{\pi(5 - 3\pi)}{3(1 + \pi)}$ & $\frac{1}{\sqrt{1 - 3\pi}}$  & $\checkmark$    & $\checkmark$ &
\\[2mm] 
   \  Sokolov & $\frac{2\pi}{1 + 3\pi}$ & $\frac{1}{\sqrt{(1 - 3\pi)(1 + \pi)}}$  & $\checkmark$  & ${\rm No}$ &
\\[2mm] 
   \  P3/P1 & $\frac{\pi(5 + 15\pi - 18\pi^2)}{3(1 + 5\pi)}$ & $\frac{(1 + \pi)^{\frac34}}{(1 + 3\pi)^{\frac14} \sqrt{1 - 3\pi}}$  & $\checkmark$  & $\checkmark$ &
\\[2.0mm] 
   \  P3/P2 & $\frac{5\pi(3\pi^2 + 7\pi - 4)}{57\pi^2 - 3\pi - 12}$ &\   $\frac{1}{(1 + \pi)^{\frac{6}{11}}(1 - \frac74 \pi)^{\frac{24}{77}}(1 - 3\pi)^{\frac{1}{2}}}$ \  &   $\checkmark$ & $\checkmark$ &
\\[2.8mm] 
   \  P3 & \  $\frac16 \pi (10 - 15\pi + 9\pi^2)$  \ & $\frac{(\pi + 1)^{\frac{3}{14}}(1-\frac34 \pi)^{\frac27}}{\sqrt{1 - 3\pi}}$  &   $\checkmark$ & No &
\\[1.8mm] 
   \  P1/P2  & $\frac{10\pi}{3(3\pi^2 + 3\pi + 2)}$ &\   $\frac{(1 + \pi)^{\frac{3}{2}}}{(4 + 3\pi)^2 \sqrt{1 - 3\pi}}$ \  &   $\checkmark$ & No &
\\[-2.5mm] 
\hspace{-20mm}  
\\[0.0mm] \hline 
\\[-3.8mm] \hline
\end{tabular}
\end{table*}
%


	\subsection{Other approximations}
\label{subsec-other-approx}
This subsection is devoted to trying to find other indicatrix functions approximating that of a relativistic Synge gas and to studying them following the same approach as in the previous subsection.\\

Firstly, we consider $\chi(\pi)$ to be the ratio of two polynomials of third and first order (P3/P1), and make it behave as the Synge indicatrix function up to second order at low temperatures and up to first order at high temperatures. With that we get
\begin{equation}
\chi(\pi) = \frac{\pi(5 + 15\pi - 18\pi^2)}{3(1 + 5\pi)} \, . \label{chi-p3/p1}
\end{equation}
Then, we can check that $\zeta>0$ and $\xi >0$, and thus both compressibility conditions (\ref{H1G}) and (\ref{H2G}) are fulfilled.

Moreover, we can determine the specific energy by using (\ref{e-psi}) and we obtain:
\begin{equation}
e(\pi) = \frac{(1 + \pi)^{\frac34}}{(1 + 3\pi)^{\frac14} \sqrt{1 - 3\pi}} \, .
\end{equation}
Then, we can see that the Taub constraint (\ref{TaubG}) also holds. 

If we compare the indicatrix function (\ref{chi-p3/p1}) of the P3/P1 EoS with that of a Synge gas, we conclude that the relative error is less than 1.25$\%$, and this EoS shows better accuracy than that of TM in the entire domain $[0, 1/3]$ (see Fig. \ref{fig-1}).

Secondly, we take $\chi(\pi)$ to be the ratio of two polynomials of the form P3/P2. Now, we can make it behave as the Synge indicatrix function up to third order at low temperatures and up to first order at high temperatures. In this case we obtain: 
\begin{eqnarray}
\chi(\pi) = \frac{5\pi(3\pi^2 + 7\pi - 4)}{57\pi^2 - 3\pi - 12} \, ,  \label{chi-p3/p2}\\[1mm]
%
%
e(\pi) = \frac{1}{(1 + \pi)^{\frac{6}{11}}(1 - \frac74 \pi)^{\frac{24}{77}}(1 - 3\pi)^{\frac{1}{2}}} \, .
\end{eqnarray}
Then, we can see that this P3/P2 EoS fulfills both compressibility conditions (\ref{H1G}) and (\ref{H2G}), and Taub's inequality (\ref{TaubG}). 

If we compare the indicatrix function (\ref{chi-p3/p2}) of the P3/P2 EoS with that of a Synge gas, we conclude that the relative error is less than 0.30$\%$ in the entire domain $[1, 1/3]$ (see Fig. \ref{fig-1}).

Finally, we look for indicatrix functions $\chi(\pi)$ that approximate the Synge EoS up to first order at both low and high temperatures. This is the case of the TM EoS, which corresponds to a ratio P2/P1. Now, we consider two new cases:  a third-order polynomial P3, and a ratio of polynomials of the form P1/P2. 

We can easily obtain the expressions for the indicatrix function $\chi(\pi)$ and the specific energy $e(\pi)$, which are listed in Table \ref{table-1}. Now, compressibility conditions (\ref{H1G}) and (\ref{H2G}) are fulfilled in both cases. Nevertheless, Taub's inequality (\ref{TaubG}) does not hold. 

Moreover, if we compare the indicatrix function of these EoS with that of a Synge gas, we conclude that the relative error is, in the entire domain $[0, 1/3]$, less than 2.90$\%$ for the P3 EoS, and less than 2.80$\%$ for the P1/P2 EoS (see Fig. \ref{fig-1}).

\section{Isentropic evolution of the Synge gas}
\label{sec-isentropic-evol}

The isentropic evolution of a nonbarotropic perfect fluid is described by a barotropic energy tensor fulfilling  a barotropic relation implicitly defined by $s (\rho, \, p) = constant$. Now, we analyze the isentropic evolution of a generic ideal gas and we particularize it to the Synge EoS and the TM approximation. The generalized Friedmann equation is stated for this case. 

	\subsection{Isentropic evolution of a generic ideal gas}
\label{subsec-GIG-isentr}

From the expression $s (\rho, \, p)$ of the specific entropy given in (\ref{s-ideal}), we find that in an isentropic evolution of a generic ideal gas we have a barotropic relation of the form:
\begin{equation} \label{rho-isentr}
\rho = K f(\pi) \, ,
\end{equation}
where $K$ is an arbitrary constant. If the function $f(\pi)$ is invertible, then (\ref{rho-isentr}) gives us the following explicit barotropic relation:
\begin{equation} \label{rel-barotropia-GIG}
p = \varphi(\rho) \equiv \rho f^{-1}(\rho / K) \, .
\end{equation}
By differentiating (\ref{rho-isentr}) and using (\ref{f-pi}) we can also characterize the barotropic relation through a differential equation:
\begin{equation} \label{rel-barotropia-diferencial-GIG}
p' \equiv \varphi'(\rho) = \chi(p/\rho) \neq \pi \, .
\end{equation}

The isentropic evolution of a classical ideal gas with adiabatic index $\gamma$ leads to a barotropic relation of the form \cite{CFS-CIG, FS-KCIG}:
\be \label{CIG-isentropic}
(\gamma-1) \rho = p + B p^{\frac{1}{\gamma}} \, , \quad B= {\rm constant} \not=0 \, .
\ee

In the case of the TM approximation, the function $f(\pi)$ is given in (\ref{e-f-TM}). Thus, by substituting it in (\ref{rho-isentr}), we obtain an implicit barotropic relation:
\begin{equation} \label{TM-barotrop}
K^2 p^3 = \rho (\rho - 3p)^4 \, .
\end{equation}

For the case of the exact Synge equation of state, we do not explicitly know $f(\pi)$ nor $\chi(\pi)$. However, if we derive (\ref{rel-barotropia-diferencial-GIG}) with respect to $\rho$ and use (\ref{chi-prima}), we can obtain a second order differential equation characterizing the barotropic relation $p=\varphi(\rho)$ leading to the isentropic evolution of a relativistic Synge gas: 
\begin{equation}
p'' \rho \pi^3 (\pi + 1) + \alpha \, p'^3 + \beta \, p'^2 + \gamma \, p' + \delta = 0 \, ,
\end{equation}
with $\alpha$, $\beta$, $\gamma$ and $\delta$ given in (\ref{chi-prima}).

	\subsection{Generic ideal gas FLRW models}
\label{subsec-FLRW-GIG}

The Friedman-Lema\^itre-Robertson-Walker universes are perfect fluid space-times with line element:
\begin{equation} \label{FLRW-metric}
ds^2 = -\dif t^2 + \frac{R^2(t)}{[1 + \frac14 k r^2]^2} (\dif r^2 + \dif \Omega^2) \, ,
\end{equation}
with $k = 0, \pm 1$, and the homogeneous energy density and pressure are given, respectively, by:
\begin{eqnarray} 
 \label{FLRW-rho}
\rho  = \frac{3 \dot{R}^2}{R^2} + \frac{3k}{R^2} + \Lambda \equiv \rho(R) \, , \\
\label{FLRW-p}
 p = - \rho - \frac{R}{3} \partial_R \rho  \equiv p(R) \, .
\end{eqnarray}

What are the {\em generalized Friedmann equations} when the energy
content is an ideal gas? The homogeneity of the hydrodynamic quantities $\rho$ and $p$ implies
that, necessarily, we have an isentropic evolution. Then, as argued in the previous subsection, a barotropic relation of the form (\ref{rel-barotropia-GIG}) holds. Then, (\ref{FLRW-p}) enables us to determine $\rho(R)$, and (\ref{FLRW-rho}) becomes a {\em Friedmann equation} for $R(t)$. 

In \cite{CFS-CIG} we have obtained this generalized Friedmann equation for the case of a classical ideal gas. Now, we analyze the case of a TM ideal gas. From the TM EoS (\ref{TM-2}) and the barotropic relation (\ref{TM-barotrop}) we obtain the evolution constraint
\be  \label{n-rho}
\rho = \rho(n) \equiv n^2(1+ \mu^2 n^{2/3}) \, ,
\ee
where $\mu \equiv 3\sqrt{3}/K$. 

On the other hand, it is known that, for the FLRW metrics, the rest-mass density is 
\begin{equation} \label{n-FLRW}
n = n_0 \left(\frac{R_0}{R}\right)^3   . 
\end{equation}
Consequently, from (\ref{n-rho}) and (\ref{n-FLRW}) we obtain the following expression for the energy density:
\begin{equation} \label{rho-R-FLRW-TM}
\rho (R) = \left[ n_0^2 \left( \frac{R_0}{R} \right)^6 + \rho_r^2 \left( \frac{R_0}{R} \right)^8 \right]^{1/2}  ,
\end{equation}
where $\rho_r \equiv 3\sqrt{3}\,n_0^{4/3}\!/K$. This expression and (\ref{FLRW-rho}) determine the generalized Friedmann equation for the TM-gas FLRW models. 

It is worth remarking that this generalized Friedmann equation was obtained by de Berredo-Peixoto {\em et al.} \cite{deBerredo} using other reasoning. In fact, they also use an equation of state equivalent to the TM EoS, which they compare with the exact Synge EoS. Here, we have used our hydrodynamic approach to show that the solutions to the Friedmann equation defined by (\ref{FLRW-rho}) and (\ref{rho-R-FLRW-TM}) model a TM gas evolving at constant entropy $s = k \ln[\rho_r/(3 \sqrt{3} n_0^{4/3})]$.

\section{Conclusions}
\label{sec-conclusions}

The hydrodynamic approach to the Synge gas presented here can be useful to look for test solutions or self-gravitating systems that model high-energy scenarios. This kind of procedure applied to other media has enabled us to analyze elsewhere the physical meaning of several families of perfect fluid solutions.    

Our study is also of conceptual interest and it has allowed us to formulate the Rainich-like theory for the Einstein-Synge solutions in subsection \ref{subsec-Rainich}. It is worth remarking that the characterization theorems presented in this subsection can be slightly changed in order to obtain the characterization of the perfect fluid solutions corresponding to the media with the other EoS considered in the paper. For example, for the TM EoS we should change condition (\ref{Si(x)}) to expression (\ref{chi-TM}) of the indicatrix function of a TM gas. 

 Several analytical approximations to the Synge EoS have shown their usefulness in numerical codes to model a relativistic gas. Here we have performed a hydrodynamic approach to recover and to analyze the Taub-Mathews EoS. We have established that the square of the speed of sound of the TM EoS and the Synge EoS differ at most by 2.36$\%$.  
 
Furthermore, our method enabled us to obtain other analytical EoS that approximate the Synge EoS. Those approximated EoS with less accuracy than the TM EoS do not fulfill Taub's inequality. Therefore, the TM equation of state acts as the limiting case. We have been able to find EoS with better accuracy than that of Taub-Mathews by reaching higher orders in the approximation.

\begin{acknowledgements}
We would like to thank the referee for his/her constructive comments. This work has been supported by the Spanish Ministerio de Ciencia, Innovaci\'on y Universidades and the Fondo Europeo de Desarrollo Regional, Projects PID2019-109753GB-C21 and PID2019-109753GB-C22 and the Generalitat Valenciana Project AICO/2020/125. S.M. acknowledges financial support from the Generalitat Valenciana (grant CIACIF/2021/028). 
\end{acknowledgements}

\bibliography{PRD}

\providecommand{\noopsort}[1]{}\providecommand{\singleletter}[1]{#1}%
\begin{thebibliography}{42}%
\makeatletter
\providecommand \@ifxundefined [1]{%
 \@ifx{#1\undefined}
}%
\providecommand \@ifnum [1]{%
 \ifnum #1\expandafter \@firstoftwo
 \else \expandafter \@secondoftwo
 \fi
}%
\providecommand \@ifx [1]{%
 \ifx #1\expandafter \@firstoftwo
 \else \expandafter \@secondoftwo
 \fi
}%
\providecommand \natexlab [1]{#1}%
\providecommand \enquote  [1]{``#1''}%
\providecommand \bibnamefont  [1]{#1}%
\providecommand \bibfnamefont [1]{#1}%
\providecommand \citenamefont [1]{#1}%
\providecommand \href@noop [0]{\@secondoftwo}%
\providecommand \href [0]{\begingroup \@sanitize@url \@href}%
\providecommand \@href[1]{\@@startlink{#1}\@@href}%
\providecommand \@@href[1]{\endgroup#1\@@endlink}%
\providecommand \@sanitize@url [0]{\catcode `\\12\catcode `\$12\catcode
  `\&12\catcode `\#12\catcode `\^12\catcode `\_12\catcode `\%12\relax}%
\providecommand \@@startlink[1]{}%
\providecommand \@@endlink[0]{}%
\providecommand \url  [0]{\begingroup\@sanitize@url \@url }%
\providecommand \@url [1]{\endgroup\@href {#1}{\urlprefix }}%
\providecommand \urlprefix  [0]{URL }%
\providecommand \Eprint [0]{\href }%
\providecommand \doibase [0]{http://dx.doi.org/}%
\providecommand \selectlanguage [0]{\@gobble}%
\providecommand \bibinfo  [0]{\@secondoftwo}%
\providecommand \bibfield  [0]{\@secondoftwo}%
\providecommand \translation [1]{[#1]}%
\providecommand \BibitemOpen [0]{}%
\providecommand \bibitemStop [0]{}%
\providecommand \bibitemNoStop [0]{.\EOS\space}%
\providecommand \EOS [0]{\spacefactor3000\relax}%
\providecommand \BibitemShut  [1]{\csname bibitem#1\endcsname}%
\let\auto@bib@innerbib\@empty
\bibitem [{\citenamefont {J$\ddot{\rm u}$ttner}(1911)}]{Juttner}%
  \BibitemOpen
  \bibfield  {author} {\bibinfo {author} {\bibfnamefont {F.}~\bibnamefont
  {J$\ddot{\rm u}$ttner}},\ }\href@noop {} {\bibfield  {journal} {\bibinfo
  {journal} {Annalen der Physik}\ }\textbf {\bibinfo {volume} {339}},\ \bibinfo
  {pages} {856} (\bibinfo {year} {1911})}\BibitemShut {NoStop}%
\bibitem [{\citenamefont {Synge}(1957)}]{Synge}%
  \BibitemOpen
  \bibfield  {author} {\bibinfo {author} {\bibfnamefont {J.~L.}\ \bibnamefont
  {Synge}},\ }\href@noop {} {\emph {\bibinfo {title} {The Relativistic Gas}}}\
  (\bibinfo  {publisher} {North-Holland, Amsterdam},\ \bibinfo {year}
  {1957})\BibitemShut {NoStop}%
\bibitem [{\citenamefont {Rezzolla}\ and\ \citenamefont
  {Zanotti}(2013)}]{Rezzolla}%
  \BibitemOpen
  \bibfield  {author} {\bibinfo {author} {\bibfnamefont {L.}~\bibnamefont
  {Rezzolla}}\ and\ \bibinfo {author} {\bibfnamefont {O.}~\bibnamefont
  {Zanotti}},\ }\href@noop {} {\emph {\bibinfo {title} {Relativistic
  hydrodynamics}}}\ (\bibinfo  {publisher} {Oxford University Press, Oxford,
  England},\ \bibinfo {year} {2013})\BibitemShut {NoStop}%
\bibitem [{\citenamefont {Mignone}\ \emph {et~al.}(2005)\citenamefont
  {Mignone}, \citenamefont {Plewa},\ and\ \citenamefont {Bodo}}]{Mignone}%
  \BibitemOpen
  \bibfield  {author} {\bibinfo {author} {\bibfnamefont {A.}~\bibnamefont
  {Mignone}}, \bibinfo {author} {\bibfnamefont {T.}~\bibnamefont {Plewa}}, \
  and\ \bibinfo {author} {\bibfnamefont {G.}~\bibnamefont {Bodo}},\ }\href@noop
  {} {\bibfield  {journal} {\bibinfo  {journal} {ApJ ss}\ }\textbf {\bibinfo
  {volume} {160}},\ \bibinfo {pages} {199} (\bibinfo {year}
  {2005})}\BibitemShut {NoStop}%
\bibitem [{\citenamefont {Chandrasekhar}(1972)}]{Chandra}%
  \BibitemOpen
  \bibfield  {author} {\bibinfo {author} {\bibfnamefont {S.}~\bibnamefont
  {Chandrasekhar}},\ }\href@noop {} {\emph {\bibinfo {title} {A limiting case
  of relativistic equilibrium in General Relativity, papers in honour of J.L.
  Synge}}}\ (\bibinfo  {publisher} {L. O' Raifeartaigh, Oxford},\ \bibinfo
  {year} {1972})\BibitemShut {NoStop}%
\bibitem [{\citenamefont {Bisnovatyi-Kogan}\ and\ \citenamefont
  {Thorne}(1970)}]{Thorne}%
  \BibitemOpen
  \bibfield  {author} {\bibinfo {author} {\bibfnamefont {G.~S.}\ \bibnamefont
  {Bisnovatyi-Kogan}}\ and\ \bibinfo {author} {\bibfnamefont {K.}~\bibnamefont
  {Thorne}},\ }\href@noop {} {\bibfield  {journal} {\bibinfo  {journal}
  {Astrophys. J.}\ }\textbf {\bibinfo {volume} {160}},\ \bibinfo {pages} {875}
  (\bibinfo {year} {1970})}\BibitemShut {NoStop}%
\bibitem [{\citenamefont {Chavanis}(2002)}]{Chavanis}%
  \BibitemOpen
  \bibfield  {author} {\bibinfo {author} {\bibfnamefont {P.~H.}\ \bibnamefont
  {Chavanis}},\ }\href@noop {} {\bibfield  {journal} {\bibinfo  {journal}
  {Astronomy and Astrophysics}\ }\textbf {\bibinfo {volume} {381}},\ \bibinfo
  {pages} {709} (\bibinfo {year} {2002})}\BibitemShut {NoStop}%
\bibitem [{\citenamefont {Krautter}\ \emph {et~al.}(1983)\citenamefont
  {Krautter}, \citenamefont {Henriksen},\ and\ \citenamefont
  {Lake}}]{Krautter}%
  \BibitemOpen
  \bibfield  {author} {\bibinfo {author} {\bibfnamefont {A.}~\bibnamefont
  {Krautter}}, \bibinfo {author} {\bibfnamefont {R.~N.}\ \bibnamefont
  {Henriksen}}, \ and\ \bibinfo {author} {\bibfnamefont {R.}~\bibnamefont
  {Lake}},\ }\href@noop {} {\bibfield  {journal} {\bibinfo  {journal}
  {Astrophys. J.}\ }\textbf {\bibinfo {volume} {269}},\ \bibinfo {pages} {81}
  (\bibinfo {year} {1983})}\BibitemShut {NoStop}%
\bibitem [{\citenamefont {Meliani}\ \emph {et~al.}(2004)\citenamefont
  {Meliani}, \citenamefont {Sauty}, \citenamefont {Tsinganos},\ and\
  \citenamefont {Vlahakis}}]{Meliani}%
  \BibitemOpen
  \bibfield  {author} {\bibinfo {author} {\bibfnamefont {Z.}~\bibnamefont
  {Meliani}}, \bibinfo {author} {\bibfnamefont {C.}~\bibnamefont {Sauty}},
  \bibinfo {author} {\bibfnamefont {K.}~\bibnamefont {Tsinganos}}, \ and\
  \bibinfo {author} {\bibfnamefont {N.}~\bibnamefont {Vlahakis}},\ }\href@noop
  {} {\bibfield  {journal} {\bibinfo  {journal} {A and A}\ }\textbf {\bibinfo
  {volume} {425}},\ \bibinfo {pages} {773} (\bibinfo {year}
  {2004})}\BibitemShut {NoStop}%
\bibitem [{\citenamefont {Lanza}\ \emph {et~al.}(1985)\citenamefont {Lanza},
  \citenamefont {Miller},\ and\ \citenamefont {Motta}}]{Lanza}%
  \BibitemOpen
  \bibfield  {author} {\bibinfo {author} {\bibfnamefont {A.}~\bibnamefont
  {Lanza}}, \bibinfo {author} {\bibfnamefont {J.~C.}\ \bibnamefont {Miller}}, \
  and\ \bibinfo {author} {\bibfnamefont {S.}~\bibnamefont {Motta}},\
  }\href@noop {} {\bibfield  {journal} {\bibinfo  {journal} {Phys. Fluids}\
  }\textbf {\bibinfo {volume} {28}},\ \bibinfo {pages} {97} (\bibinfo {year}
  {1985})}\BibitemShut {NoStop}%
\bibitem [{\citenamefont {Scheck}\ \emph {et~al.}(2002)\citenamefont {Scheck},
  \citenamefont {Aloy}, \citenamefont {Mart\'{\i}}, \citenamefont {G\'omez},\
  and\ \citenamefont {M$\ddot{\rm u}$ler}}]{Scheck-Aloy}%
  \BibitemOpen
  \bibfield  {author} {\bibinfo {author} {\bibfnamefont {L.}~\bibnamefont
  {Scheck}}, \bibinfo {author} {\bibfnamefont {M.~A.}\ \bibnamefont {Aloy}},
  \bibinfo {author} {\bibfnamefont {J.~M.}\ \bibnamefont {Mart\'{\i}}},
  \bibinfo {author} {\bibfnamefont {J.~L.}\ \bibnamefont {G\'omez}}, \ and\
  \bibinfo {author} {\bibfnamefont {E.}~\bibnamefont {M$\ddot{\rm u}$ler}},\
  }\href@noop {} {\bibfield  {journal} {\bibinfo  {journal} {Mon. Not. R.
  Astron. Soc.}\ }\textbf {\bibinfo {volume} {311}},\ \bibinfo {pages} {615}
  (\bibinfo {year} {2002})}\BibitemShut {NoStop}%
\bibitem [{\citenamefont {Perucho}\ and\ \citenamefont
  {Mart\'{\i}}(2007)}]{Perucho_2007}%
  \BibitemOpen
  \bibfield  {author} {\bibinfo {author} {\bibfnamefont {M.}~\bibnamefont
  {Perucho}}\ and\ \bibinfo {author} {\bibfnamefont {J.~M.}\ \bibnamefont
  {Mart\'{\i}}},\ }\href@noop {} {\bibfield  {journal} {\bibinfo  {journal}
  {Mon. Not. R. Astron. Soc.}\ }\textbf {\bibinfo {volume} {382}},\ \bibinfo
  {pages} {626} (\bibinfo {year} {2007})}\BibitemShut {NoStop}%
\bibitem [{\citenamefont {Choi}\ and\ \citenamefont {Wiita}(2010)}]{Choi}%
  \BibitemOpen
  \bibfield  {author} {\bibinfo {author} {\bibfnamefont {E.}~\bibnamefont
  {Choi}}\ and\ \bibinfo {author} {\bibfnamefont {P.~J.}\ \bibnamefont
  {Wiita}},\ }\href@noop {} {\bibfield  {journal} {\bibinfo  {journal}
  {Astrophisical Journal SS}\ }\textbf {\bibinfo {volume} {191}},\ \bibinfo
  {pages} {113} (\bibinfo {year} {2010})}\BibitemShut {NoStop}%
\bibitem [{\citenamefont {Perucho}\ \emph {et~al.}(2014)\citenamefont
  {Perucho}, \citenamefont {Mart\'{\i}}, \citenamefont {Laing},\ and\
  \citenamefont {Hardee}}]{Perucho_2014}%
  \BibitemOpen
  \bibfield  {author} {\bibinfo {author} {\bibfnamefont {M.}~\bibnamefont
  {Perucho}}, \bibinfo {author} {\bibfnamefont {J.~M.}\ \bibnamefont
  {Mart\'{\i}}}, \bibinfo {author} {\bibfnamefont {R.~A.}\ \bibnamefont
  {Laing}}, \ and\ \bibinfo {author} {\bibfnamefont {P.~E.}\ \bibnamefont
  {Hardee}},\ }\href@noop {} {\bibfield  {journal} {\bibinfo  {journal} {Mon.
  Not. R. Astron. Soc.}\ }\textbf {\bibinfo {volume} {441}},\ \bibinfo {pages}
  {1488} (\bibinfo {year} {2014})}\BibitemShut {NoStop}%
\bibitem [{\citenamefont {Perucho}\ \emph {et~al.}(2019)\citenamefont
  {Perucho}, \citenamefont {Mart\'{\i}},\ and\ \citenamefont
  {Quilis}}]{Perucho_2019}%
  \BibitemOpen
  \bibfield  {author} {\bibinfo {author} {\bibfnamefont {M.}~\bibnamefont
  {Perucho}}, \bibinfo {author} {\bibfnamefont {J.~M.}\ \bibnamefont
  {Mart\'{\i}}}, \ and\ \bibinfo {author} {\bibfnamefont {V.}~\bibnamefont
  {Quilis}},\ }\href@noop {} {\bibfield  {journal} {\bibinfo  {journal} {Mon.
  Not. R. Astron. Soc.}\ }\textbf {\bibinfo {volume} {482}},\ \bibinfo {pages}
  {3718} (\bibinfo {year} {2019})}\BibitemShut {NoStop}%
\bibitem [{\citenamefont {Angl\'es-Castillo}\ \emph {et~al.}(2021)\citenamefont
  {Angl\'es-Castillo}, \citenamefont {Perucho}, \citenamefont {Mart\'{\i}},\
  and\ \citenamefont {Laing}}]{Perucho_2021}%
  \BibitemOpen
  \bibfield  {author} {\bibinfo {author} {\bibfnamefont {A.}~\bibnamefont
  {Angl\'es-Castillo}}, \bibinfo {author} {\bibfnamefont {M.}~\bibnamefont
  {Perucho}}, \bibinfo {author} {\bibfnamefont {J.~M.}\ \bibnamefont
  {Mart\'{\i}}}, \ and\ \bibinfo {author} {\bibfnamefont {R.~A.}\ \bibnamefont
  {Laing}},\ }\href@noop {} {\bibfield  {journal} {\bibinfo  {journal} {Mon.
  Not. R. Astron. Soc.}\ }\textbf {\bibinfo {volume} {500}},\ \bibinfo {pages}
  {1512} (\bibinfo {year} {2021})}\BibitemShut {NoStop}%
\bibitem [{\citenamefont {Perucho}\ \emph {et~al.}(2022)\citenamefont
  {Perucho}, \citenamefont {Mart\'{\i}},\ and\ \citenamefont
  {Quilis}}]{Perucho_2022}%
  \BibitemOpen
  \bibfield  {author} {\bibinfo {author} {\bibfnamefont {M.}~\bibnamefont
  {Perucho}}, \bibinfo {author} {\bibfnamefont {J.~M.}\ \bibnamefont
  {Mart\'{\i}}}, \ and\ \bibinfo {author} {\bibfnamefont {V.}~\bibnamefont
  {Quilis}},\ }\href@noop {} {\bibfield  {journal} {\bibinfo  {journal} {Mon.
  Not. R. Astron. Soc.}\ }\textbf {\bibinfo {volume} {510}},\ \bibinfo {pages}
  {2084} (\bibinfo {year} {2022})}\BibitemShut {NoStop}%
\bibitem [{\citenamefont {Szekeres}\ and\ \citenamefont
  {Barnes}(1979)}]{Szekeres-Barnes}%
  \BibitemOpen
  \bibfield  {author} {\bibinfo {author} {\bibfnamefont {P.}~\bibnamefont
  {Szekeres}}\ and\ \bibinfo {author} {\bibfnamefont {A.~S.}\ \bibnamefont
  {Barnes}},\ }\href@noop {} {\bibfield  {journal} {\bibinfo  {journal} {Mon.
  Not. R. astr. Soc.}\ }\textbf {\bibinfo {volume} {189}},\ \bibinfo {pages}
  {767} (\bibinfo {year} {1979})}\BibitemShut {NoStop}%
\bibitem [{\citenamefont {de~Berredo-Peixoto}\ \emph
  {et~al.}(2005)\citenamefont {de~Berredo-Peixoto}, \citenamefont {Shaphiro},\
  and\ \citenamefont {Sobreira}}]{deBerredo}%
  \BibitemOpen
  \bibfield  {author} {\bibinfo {author} {\bibfnamefont {G.}~\bibnamefont
  {de~Berredo-Peixoto}}, \bibinfo {author} {\bibfnamefont {I.~L.}\ \bibnamefont
  {Shaphiro}}, \ and\ \bibinfo {author} {\bibfnamefont {F.}~\bibnamefont
  {Sobreira}},\ }\href@noop {} {\bibfield  {journal} {\bibinfo  {journal}
  {Phys. Lett. A}\ }\textbf {\bibinfo {volume} {20}},\ \bibinfo {pages} {2723}
  (\bibinfo {year} {2005})}\BibitemShut {NoStop}%
\bibitem [{\citenamefont {Pordeus-daSilva}\ \emph {et~al.}(2019)\citenamefont
  {Pordeus-daSilva}, \citenamefont {Batista},\ and\ \citenamefont
  {Medeiros}}]{Silva-2019}%
  \BibitemOpen
  \bibfield  {author} {\bibinfo {author} {\bibfnamefont {G.}~\bibnamefont
  {Pordeus-daSilva}}, \bibinfo {author} {\bibfnamefont {R.~C.}\ \bibnamefont
  {Batista}}, \ and\ \bibinfo {author} {\bibfnamefont {L.~G.}\ \bibnamefont
  {Medeiros}},\ }\href@noop {} {\bibfield  {journal} {\bibinfo  {journal}
  {JCAP}\ }\textbf {\bibinfo {volume} {06}},\ \bibinfo {pages} {043} (\bibinfo
  {year} {2019})}\BibitemShut {NoStop}%
\bibitem [{\citenamefont {Eckart}(1940)}]{Eckart}%
  \BibitemOpen
  \bibfield  {author} {\bibinfo {author} {\bibfnamefont {C.}~\bibnamefont
  {Eckart}},\ }\href@noop {} {\bibfield  {journal} {\bibinfo  {journal} {Phys.
  Rev.}\ }\textbf {\bibinfo {volume} {58}},\ \bibinfo {pages} {919} (\bibinfo
  {year} {1940})}\BibitemShut {NoStop}%
\bibitem [{\citenamefont {Coll}\ and\ \citenamefont
  {Ferrando}(1989)}]{Coll-Ferrando-termo}%
  \BibitemOpen
  \bibfield  {author} {\bibinfo {author} {\bibfnamefont {B.}~\bibnamefont
  {Coll}}\ and\ \bibinfo {author} {\bibfnamefont {J.~J.}\ \bibnamefont
  {Ferrando}},\ }\href@noop {} {\bibfield  {journal} {\bibinfo  {journal} {J.
  Math. Phys.}\ }\textbf {\bibinfo {volume} {30}},\ \bibinfo {pages} {2918}
  (\bibinfo {year} {1989})}\BibitemShut {NoStop}%
\bibitem [{\citenamefont {Coll}\ \emph {et~al.}(2017)\citenamefont {Coll},
  \citenamefont {Ferrando},\ and\ \citenamefont {S\'aez}}]{CFS-LTE}%
  \BibitemOpen
  \bibfield  {author} {\bibinfo {author} {\bibfnamefont {B.}~\bibnamefont
  {Coll}}, \bibinfo {author} {\bibfnamefont {J.~J.}\ \bibnamefont {Ferrando}},
  \ and\ \bibinfo {author} {\bibfnamefont {J.~A.}\ \bibnamefont {S\'aez}},\
  }\href@noop {} {\bibfield  {journal} {\bibinfo  {journal} {Gen. Relativ.
  Gravit.}\ }\textbf {\bibinfo {volume} {49}},\ \bibinfo {pages} {66} (\bibinfo
  {year} {2017})}\BibitemShut {NoStop}%
\bibitem [{\citenamefont {Coll}\ \emph
  {et~al.}(2020{\natexlab{a}})\citenamefont {Coll}, \citenamefont {Ferrando},\
  and\ \citenamefont {S\'aez}}]{CFS-CC}%
  \BibitemOpen
  \bibfield  {author} {\bibinfo {author} {\bibfnamefont {B.}~\bibnamefont
  {Coll}}, \bibinfo {author} {\bibfnamefont {J.~J.}\ \bibnamefont {Ferrando}},
  \ and\ \bibinfo {author} {\bibfnamefont {J.~A.}\ \bibnamefont {S\'aez}},\
  }\href@noop {} {\bibfield  {journal} {\bibinfo  {journal} {Phys. Rev. D}\
  }\textbf {\bibinfo {volume} {101}},\ \bibinfo {pages} {064058} (\bibinfo
  {year} {2020}{\natexlab{a}})}\BibitemShut {NoStop}%
\bibitem [{\citenamefont {Coll}\ and\ \citenamefont
  {Ferrando}(2005)}]{CF-Stephani}%
  \BibitemOpen
  \bibfield  {author} {\bibinfo {author} {\bibfnamefont {B.}~\bibnamefont
  {Coll}}\ and\ \bibinfo {author} {\bibfnamefont {J.~J.}\ \bibnamefont
  {Ferrando}},\ }\href@noop {} {\bibfield  {journal} {\bibinfo  {journal} {Gen.
  Relativ. Gravit.}\ }\textbf {\bibinfo {volume} {37}},\ \bibinfo {pages} {557}
  (\bibinfo {year} {2005})}\BibitemShut {NoStop}%
\bibitem [{\citenamefont {Coll}\ \emph
  {et~al.}(2019{\natexlab{a}})\citenamefont {Coll}, \citenamefont {Ferrando},\
  and\ \citenamefont {S\'aez}}]{CFS-CIG}%
  \BibitemOpen
  \bibfield  {author} {\bibinfo {author} {\bibfnamefont {B.}~\bibnamefont
  {Coll}}, \bibinfo {author} {\bibfnamefont {J.~J.}\ \bibnamefont {Ferrando}},
  \ and\ \bibinfo {author} {\bibfnamefont {J.~A.}\ \bibnamefont {S\'aez}},\
  }\href@noop {} {\bibfield  {journal} {\bibinfo  {journal} {Phys. Rev. D}\
  }\textbf {\bibinfo {volume} {99}},\ \bibinfo {pages} {084035} (\bibinfo
  {year} {2019}{\natexlab{a}})}\BibitemShut {NoStop}%
\bibitem [{\citenamefont {Ferrando}\ and\ \citenamefont
  {S\'aez}(2018)}]{FS-SS}%
  \BibitemOpen
  \bibfield  {author} {\bibinfo {author} {\bibfnamefont {J.~J.}\ \bibnamefont
  {Ferrando}}\ and\ \bibinfo {author} {\bibfnamefont {J.~A.}\ \bibnamefont
  {S\'aez}},\ }\href@noop {} {\bibfield  {journal} {\bibinfo  {journal} {Phys.
  Rev. D}\ }\textbf {\bibinfo {volume} {97}},\ \bibinfo {pages} {044026}
  (\bibinfo {year} {2018})}\BibitemShut {NoStop}%
\bibitem [{\citenamefont {Coll}\ \emph
  {et~al.}(2019{\natexlab{b}})\citenamefont {Coll}, \citenamefont {Ferrando},\
  and\ \citenamefont {S\'aez}}]{CFS-PSS}%
  \BibitemOpen
  \bibfield  {author} {\bibinfo {author} {\bibfnamefont {B.}~\bibnamefont
  {Coll}}, \bibinfo {author} {\bibfnamefont {J.~J.}\ \bibnamefont {Ferrando}},
  \ and\ \bibinfo {author} {\bibfnamefont {J.~A.}\ \bibnamefont {S\'aez}},\
  }\href@noop {} {\bibfield  {journal} {\bibinfo  {journal} {Class. Quantum
  Grav.}\ }\textbf {\bibinfo {volume} {36}},\ \bibinfo {pages} {175004}
  (\bibinfo {year} {2019}{\natexlab{b}})}\BibitemShut {NoStop}%
\bibitem [{\citenamefont {Coll}\ \emph
  {et~al.}(2020{\natexlab{b}})\citenamefont {Coll}, \citenamefont {Ferrando},\
  and\ \citenamefont {S\'aez}}]{CFS-RSS}%
  \BibitemOpen
  \bibfield  {author} {\bibinfo {author} {\bibfnamefont {B.}~\bibnamefont
  {Coll}}, \bibinfo {author} {\bibfnamefont {J.~J.}\ \bibnamefont {Ferrando}},
  \ and\ \bibinfo {author} {\bibfnamefont {J.~A.}\ \bibnamefont {S\'aez}},\
  }\href@noop {} {\bibfield  {journal} {\bibinfo  {journal} {Class. Quantum
  Grav.}\ }\textbf {\bibinfo {volume} {37}},\ \bibinfo {pages} {185005}
  (\bibinfo {year} {2020}{\natexlab{b}})}\BibitemShut {NoStop}%
\bibitem [{\citenamefont {Ferrando}\ and\ \citenamefont
  {Mengual}(2021{\natexlab{a}})}]{FM-Tmodels}%
  \BibitemOpen
  \bibfield  {author} {\bibinfo {author} {\bibfnamefont {J.~J.}\ \bibnamefont
  {Ferrando}}\ and\ \bibinfo {author} {\bibfnamefont {S.}~\bibnamefont
  {Mengual}},\ }\href@noop {} {\bibfield  {journal} {\bibinfo  {journal} {Phys.
  Rev. D}\ }\textbf {\bibinfo {volume} {104}},\ \bibinfo {pages} {024038}
  (\bibinfo {year} {2021}{\natexlab{a}})}\BibitemShut {NoStop}%
\bibitem [{\citenamefont {Ferrando}\ and\ \citenamefont
  {Mengual}(2021{\natexlab{b}})}]{FM-Tmodels2}%
  \BibitemOpen
  \bibfield  {author} {\bibinfo {author} {\bibfnamefont {J.~J.}\ \bibnamefont
  {Ferrando}}\ and\ \bibinfo {author} {\bibfnamefont {S.}~\bibnamefont
  {Mengual}},\ }\href@noop {} {\bibfield  {journal} {\bibinfo  {journal} {Phys.
  Rev. D}\ }\textbf {\bibinfo {volume} {104}},\ \bibinfo {pages} {064029}
  (\bibinfo {year} {2021}{\natexlab{b}})}\BibitemShut {NoStop}%
\bibitem [{\citenamefont {Mengual}\ and\ \citenamefont
  {Ferrando}(2022)}]{MF-LT}%
  \BibitemOpen
  \bibfield  {author} {\bibinfo {author} {\bibfnamefont {S.}~\bibnamefont
  {Mengual}}\ and\ \bibinfo {author} {\bibfnamefont {J.~J.}\ \bibnamefont
  {Ferrando}},\ }\href@noop {} {\bibfield  {journal} {\bibinfo  {journal}
  {Phys. Rev. D}\ }\textbf {\bibinfo {volume} {105}},\ \bibinfo {pages}
  {124019} (\bibinfo {year} {2022})}\BibitemShut {NoStop}%
\bibitem [{\citenamefont {Taub}(1948)}]{Taub}%
  \BibitemOpen
  \bibfield  {author} {\bibinfo {author} {\bibfnamefont {A.~H.}\ \bibnamefont
  {Taub}},\ }\href@noop {} {\bibfield  {journal} {\bibinfo  {journal} {Phys.
  Rev.}\ }\textbf {\bibinfo {volume} {74}},\ \bibinfo {pages} {328} (\bibinfo
  {year} {1948})}\BibitemShut {NoStop}%
\bibitem [{\citenamefont {Mathews}(1971)}]{Mathews}%
  \BibitemOpen
  \bibfield  {author} {\bibinfo {author} {\bibfnamefont {W.~G.}\ \bibnamefont
  {Mathews}},\ }\href@noop {} {\bibfield  {journal} {\bibinfo  {journal} {ApJ}\
  }\textbf {\bibinfo {volume} {165}},\ \bibinfo {pages} {147} (\bibinfo {year}
  {1971})}\BibitemShut {NoStop}%
\bibitem [{\citenamefont {Mignone}\ and\ \citenamefont
  {McKinney}(2007)}]{Mignone-2007}%
  \BibitemOpen
  \bibfield  {author} {\bibinfo {author} {\bibfnamefont {A.}~\bibnamefont
  {Mignone}}\ and\ \bibinfo {author} {\bibfnamefont {J.~C.}\ \bibnamefont
  {McKinney}},\ }\href@noop {} {\bibfield  {journal} {\bibinfo  {journal} {Mon.
  Not. R. Astron. Soc.}\ }\textbf {\bibinfo {volume} {387}},\ \bibinfo {pages}
  {1118} (\bibinfo {year} {2007})}\BibitemShut {NoStop}%
\bibitem [{\citenamefont {Sokolov}\ \emph {et~al.}(2001)\citenamefont
  {Sokolov}, \citenamefont {Zhang},\ and\ \citenamefont {Sakai}}]{Sokolov}%
  \BibitemOpen
  \bibfield  {author} {\bibinfo {author} {\bibfnamefont {I.~V.}\ \bibnamefont
  {Sokolov}}, \bibinfo {author} {\bibfnamefont {H.-M.}\ \bibnamefont {Zhang}},
  \ and\ \bibinfo {author} {\bibfnamefont {J.~I.}\ \bibnamefont {Sakai}},\
  }\href@noop {} {\bibfield  {journal} {\bibinfo  {journal} {Journal of
  Computational Physics}\ }\textbf {\bibinfo {volume} {172}},\ \bibinfo {pages}
  {209} (\bibinfo {year} {2001})}\BibitemShut {NoStop}%
\bibitem [{\citenamefont {Pleba\'nski}(1964)}]{Plebanski}%
  \BibitemOpen
  \bibfield  {author} {\bibinfo {author} {\bibfnamefont {J.}~\bibnamefont
  {Pleba\'nski}},\ }\href@noop {} {\bibfield  {journal} {\bibinfo  {journal}
  {Acta Phys. Pol.}\ }\textbf {\bibinfo {volume} {26}},\ \bibinfo {pages} {963}
  (\bibinfo {year} {1964})}\BibitemShut {NoStop}%
\bibitem [{\citenamefont {Israel}(1960)}]{Israel}%
  \BibitemOpen
  \bibfield  {author} {\bibinfo {author} {\bibfnamefont {W.}~\bibnamefont
  {Israel}},\ }\href@noop {} {\bibfield  {journal} {\bibinfo  {journal} {Proc.
  R. Soc. London}\ }\textbf {\bibinfo {volume} {259}},\ \bibinfo {pages} {129}
  (\bibinfo {year} {1960})}\BibitemShut {NoStop}%
\bibitem [{\citenamefont {Lichnerowicz}(1966)}]{Lichnero-1}%
  \BibitemOpen
  \bibfield  {author} {\bibinfo {author} {\bibfnamefont {A.}~\bibnamefont
  {Lichnerowicz}},\ }\href@noop {} {\bibfield  {journal} {\bibinfo  {journal}
  {Ann. Inst. Henri Poincar\'e}\ }\textbf {\bibinfo {volume} {5}},\ \bibinfo
  {pages} {37} (\bibinfo {year} {1966})}\BibitemShut {NoStop}%
\bibitem [{\citenamefont {Rainich}(1925)}]{Rainich}%
  \BibitemOpen
  \bibfield  {author} {\bibinfo {author} {\bibfnamefont {G.~Y.}\ \bibnamefont
  {Rainich}},\ }\href@noop {} {\bibfield  {journal} {\bibinfo  {journal}
  {Trans. Math. Soc.}\ }\textbf {\bibinfo {volume} {27}},\ \bibinfo {pages}
  {106} (\bibinfo {year} {1925})}\BibitemShut {NoStop}%
\bibitem [{\citenamefont {Anile}(1989)}]{Anile}%
  \BibitemOpen
  \bibfield  {author} {\bibinfo {author} {\bibfnamefont {A.~M.}\ \bibnamefont
  {Anile}},\ }\href@noop {} {\emph {\bibinfo {title} {Relativistic fluids and
  magneto-fluids}}}\ (\bibinfo  {publisher} {Cambridge University Press,
  Cambridge, England},\ \bibinfo {year} {1989})\BibitemShut {NoStop}%
\bibitem [{\citenamefont {Ferrando}\ and\ \citenamefont
  {S\'aez}(2019)}]{FS-KCIG}%
  \BibitemOpen
  \bibfield  {author} {\bibinfo {author} {\bibfnamefont {J.~J.}\ \bibnamefont
  {Ferrando}}\ and\ \bibinfo {author} {\bibfnamefont {J.~A.}\ \bibnamefont
  {S\'aez}},\ }\href@noop {} {\bibfield  {journal} {\bibinfo  {journal} {Class.
  Quantum Grav.}\ }\textbf {\bibinfo {volume} {36}},\ \bibinfo {pages} {215008}
  (\bibinfo {year} {2019})}\BibitemShut {NoStop}%
\end{thebibliography}%

\end{document}